\documentclass{aa}
\usepackage[T1]{fontenc}
\usepackage{multirow}
\usepackage{amstext}
\usepackage{amsmath}
\usepackage{graphicx}
\usepackage{tabularx}
\usepackage{siunitx}
\usepackage[varg]{txfonts}  
\bibpunct{(}{)}{;}{a}{}{,}  
\usepackage[colorlinks, citecolor=blue, linkcolor=blue, urlcolor  = blue]{hyperref}
\usepackage{placeins} 

\usepackage[T1]{fontenc}
\usepackage[utf8]{inputenc}

\makeatletter

\iftrue
    \usepackage{twoopt}
    \usepackage{xstring}
    \newcommand{\adshref}[2]{\StrRight{#1}{19}[\adsid]\href{http://adsabs.harvard.edu/abs/\adsid}{#2}}
    
    \let\orgcitep\citep
    \let\orgcitet\citet
    \let\orgcitealt\citealt
    \let\orgcitealp\citealp
    \let\orgciteauthor\citeauthor

    \renewcommandtwoopt{\cite}[3][][]{\adshref{#3}
        {\def\hyper@linkstart##1##2{}
        \let\hyper@linkend\@empty\orgcitet[#1][#2]{#3}}}
    \renewcommandtwoopt{\citep}[3][][]{\adshref{#3}
        {\def\hyper@linkstart##1##2{}
        \let\hyper@linkend\@empty\orgcitep[#1][#2]{#3}}}
    \renewcommandtwoopt{\citet}[3][][]{\adshref{#3}
        {\def\hyper@linkstart##1##2{}
        \let\hyper@linkend\@empty\orgcitet[#1][#2]{#3}}}
    \renewcommandtwoopt{\citealt}[3][][]{\adshref{#3}
        {\def\hyper@linkstart##1##2{}
        \let\hyper@linkend\@empty\orgcitealt[#1][#2]{#3}}}
    \renewcommandtwoopt{\citealp}[3][][]{\adshref{#3}
        {\def\hyper@linkstart##1##2{}
        \let\hyper@linkend\@empty\orgcitealp[#1][#2]{#3}}}
    \renewcommandtwoopt{\citeauthor}[3][][]{\adshref{#3}
        {\def\hyper@linkstart##1##2{}
        \let\hyper@linkend\@empty\orgciteauthor[#1][#2]{#3}}}
    \newcommandtwoopt{\citeyearads}[3][][]
        {\href{http://adsabs.harvard.edu/abs/#3}
        {\def\hyper@linkstart##1##2{}
        \let\hyper@linkend\@empty\citeyear[#1][#2]{#3}}}
   \iftrue
        \renewcommandtwoopt{\cite}[3][][]{\adshref{#3}{\orgcitet[#1][#2]{#3}}}
        \renewcommandtwoopt{\citep}[3][][]{\adshref{#3}{\orgcitep[#1][#2]{#3}}}
        \renewcommandtwoopt{\citet}[3][][]{\adshref{#3}{\orgcitet[#1][#2]{#3}}}
        \renewcommandtwoopt{\citealt}[3][][]{\adshref{#3}{\orgcitealt[#1][#2]{#3}}}
        \renewcommandtwoopt{\citealp}[3][][]{\adshref{#3}{\orgcitealp[#1][#2]{#3}}}
    \fi
\fi

\def\instrefs#1{{\def\scsep{\def\scsep{,}}\@for\w:=#1\do{\scsep\ref{inst:\w}}}}
\renewcommand{\inst}[1]{\unskip$^{\instrefs{#1}}$}

\usepackage{xspace}

\renewcommand*\aa@pageof{, page \thepage{} of \pageref*{LastPage}} 

\usepackage{academicons}
\usepackage{xcolor}
\newcommand{\orcid}[1]{\href{https://orcid.org/#1}{\textcolor[HTML]{A6CE39}{\aiOrcid}}}

\usepackage{scalerel}
\usepackage{tikz}
\usetikzlibrary{svg.path}
\definecolor{orcidlogocol}{HTML}{A6CE39}
\tikzset{
  orcidlogo/.pic={
    \fill[orcidlogocol] svg{M256,128c0,70.7-57.3,128-128,128C57.3,256,0,198.7,0,128C0,57.3,57.3,0,128,0C198.7,0,256,57.3,256,128z};
    \fill[white] svg{M86.3,186.2H70.9V79.1h15.4v48.4V186.2z}
                 svg{M108.9,79.1h41.6c39.6,0,57,28.3,57,53.6c0,27.5-21.5,53.6-56.8,53.6h-41.8V79.1z M124.3,172.4h24.5c34.9,0,42.9-26.5,42.9-39.7c0-21.5-13.7-39.7-43.7-39.7h-23.7V172.4z}
                 svg{M88.7,56.8c0,5.5-4.5,10.1-10.1,10.1c-5.6,0-10.1-4.6-10.1-10.1c0-5.6,4.5-10.1,10.1-10.1C84.2,46.7,88.7,51.3,88.7,56.8z};
  }
}
\newcommand\orcidicon[1]{\href{https://orcid.org/#1}{\mbox{\scalerel*{
\begin{tikzpicture}[yscale=-1,transform shape]
\pic{orcidlogo};
\end{tikzpicture}
}{|}}}}
\onecolumn

\begin{document}

\lefthyphenmin=3

\title{RedDots: Planetary masses in the GJ1061 system from planet-planet interaction
\thanks{Based on observations carried out at the European Southern Observatory under ESO programme 072.C-0488(E), 0101.C-0516(A), 198.C-0838(A), 1102.C-0339(A), and 183.C-0437(A)}} 

\titlerunning{RedDots: Planetary masses in the GJ\,1061 system from planet-planet interaction}
\authorrunning{S. Dreizler et al.}

\author{
    S.~Dreizler\inst{iag}
    \and S.\,V.~Jeffers\inst{tls}
    \and F.~Liebing\inst{pa}
    \and P.~Gorrini\inst{iag}
    \and C.\,A.~Haswell\inst{open}
    \and E.~Gaidos\inst{hawaii,vienna}
    \and J.~R.~Barnes\inst{open}
    \and F.~Del~Sordo\inst{ice,ieec,inaf}
    \and H.\,R.\,A.~Jones\inst{car}
    \and E.~Rodríguez\inst{iac}
    \and Y.~Tsapras\inst{arizah}
}

\institute{
    \label{inst:iag}Institut f\"ur Astrophysik, Georg-August-Universit\"at, Friedrich-Hund-Platz 1, 37077 G\"ottingen, Germany\\
    \email{dreizler@astro.physik.uni-goettingen.de}
    \and \label{inst:tls} Th\"uringer Landessternwarte Tautenburg, Sternwarte 5, D-07778 Tautenburg, Germany 
    \and \label{inst:pa} Private Astronomer
    \and \label{inst:open} Department of Physics and Astronomy, The Open University, Walton Hall, Milton Keynes MK7 6AA, U.K. 
    \and \label{inst:hawaii}Department of Earth Sciences, University of Hawai'i at M\"{a}noa, 1680 East-West Road, Honolulu, HI 96822, USA
    \and \label{inst:vienna}Institut f\"ur Astrophysik, Universit"at Wien, T\"{u}rkenschanzstra{\ss}e 17, 1180 Wien, Austria 
    \and \label{inst:ice} Institute of Space Sciences (ICE-CSIC), Campus UAB, Carrer de Can Magrans s/n, 08193, Barcelona, Spain
    \and \label{inst:ieec} Institut d’Estudis Espacials de Catalunya (IEEC), 08034 Barcelona, Spain
    \and \label{inst:inaf} INAF, Osservatorio Astrofisico di Catania, via Santa Sofia, 78 Catania, Italy
    \and \label{inst:car} Physics, Astronomy and Mathematics, University of Hertfordshire, College Lane, Hatfield AL10 9AB, UK
    \and \label{inst:iac} Instituto de Astrofísica de Andalucía (IAA-CSIC); Glorieta de la Astronomía, s/n; E-18008 Granada, Spain
    \and \label{inst:arizah}Zentrum f{\"u}r Astronomie der Universit{\"a}t Heidelberg, Astronomisches Rechen-Institut, M{\"o}nchhofstr. 12-14, 69120 Heidelberg, Germany
}

\abstract{GJ\,1061 is a very nearby M star hosting three low-mass temperate planets detected from radial velocity variations. The close to 4:2:1 period commensurability of the planets, the available long-term monitoring of the system and new very high-precision radial velocity measurements from ESPRESSO enable the determination of masses from the planet-planet interaction. Using nested sampling we derived parameter distributions for a co-planar configuration. The three planets (M$_{\text b}$=$1.07\pm 0.11M_\oplus$, P$_{\text b}$=$3.2073\pm 0.0003$\,d, M$_{\text c}$=$1.76\pm 0.13M_\oplus$, P$_{\text c}$=$6.6821\pm 0.0008$\,d, M$_{\text d}$=$1.55\pm 0.17M_\oplus$, P$_{\text d}$=$13.066\pm 0.002$\,d) are potentially all rocky with equilibrium temperatures between 360\,K and 240\,K. This makes the GJ\,1061 system one of the prime targets for future ground or space based instruments suitable for a direct detection of the planetary atmospheres.}

\keywords{planets and satellites: detection -- fundamental parameters -- individual: GJ\,1061 bcd -- terrestrial planets / stars: planetary systems -- individual: GJ\,1061}

\maketitle
\sloppy

\section{Introduction}

One of the greatest achievements in astrophysics in recent times is the detection of over 5000 planets orbiting stars beyond our Solar System.  Since the first detection of exoplanets, \citep{1989Natur.339...38L,1995Natur.378..355M} the vast majority of planets have been discovered using the radial velocity (RV) and transit techniques.  Of particular interest is the characterisation of the atmospheres and internal composition of rocky-Earth mass planets orbiting in the liquid water habitable zones of their host stars.  In the Habitable Worlds Catalog (HWC)\footnote{\url{phl.upr.edu/hwc}}, however, there are only 23 detections of potential rocky ($0.5\,{\rm R_\odot} < R_{\rm p} \leq 1.6\,{\rm R_\odot}$ or $0.1\,{\rm M_\odot} < M_{\rm p, min} \leq 3\,{\rm M_\odot}$) planets that orbit in their host star's conservative habitable zone \citep{2014ApJ...787L..29K}. Among these, only the planets in the TRAPPIST-1-system have been robustly determined to be rocky from accurate mass and radius measurements. Nine out of the 23 potentially habitable worlds were detected by the radial velocity method, the other 14 by the transit method.

The lack of robustly determined planetary mass values is primarily because the RV technique can only provide a lower limit on a planet's mass unless the inclination is known from other measurements, such as via planetary transits. However, the probability that a planet transits is low in general. A planet in the habitable zone of a late M dwarf like GJ\,1061 has a transit probability of about 1\%. Moreover, the majority of known transiting systems are at larger distances to our Sun, with relatively faint host star apparent brightnesses, making characterisation of their atmospheres more challenging.
The transit planets in the HWC for example have a mean distance of 105\,pc while the RV planets have a mean distance of only 3.4\,pc. The latter are therefore prime targets for direct spectroscopic investigations of planet atmospheres in the era of extremely large telescopes with instruments like the ArmazoNes high Dispersion Echelle Spectrograph \citep{2023arXiv231117075P} or the Planetary Camera Spectrograph \citep{2021Msngr.182...38K} and with space missions such as the Habitable Worlds Observatory \citep{NAP26141} or the complementary Large Interferometer For Exoplanets mission \citep{2022A&A...664A..21Q,2022ExA....54.1197Q}.

In multi-planet systems, the strength of the gravitational planet-planet interactions depends on the masses and mutual inclinations of the planets, and opens up the possibility to obtaining more precise planetary parameters. For dynamically compact systems, even small deviations from Keplerian orbits can become detectable within just a few years for close orbiting planets, especially with the advent of ultra-high precision spectrographs such as ESPRESSO \citep{2021A&A...645A..96P}. An early example of the application of planet-planet interactions as probe of the planetary masses is GJ\,876. It is among the first detected exoplanets, \citep{1998ApJ...505L.147M,1998A&A...338L..67D} and the dynamical interaction between the two Jupiter-mass planets in a 2:1 mean motion resonance was first modelled by \citet{2005ApJ...634..625R} and later refined by \citet{2010ApJ...719..890R} and \citet{2018A&A...609A.117T}. Another interesting case is the GJ\,581 system. The first three planets were detected by \cite{Bonfils2005} and \cite{Udry2007}. \citet{Vogt2010} announced GJ\,581 as a possible six-planet system, but stellar activity has been identified as the likely cause of three of the signals \citep{Baluev2013,Robertson2014,Hatzes2016}. The planet-planet interaction of the confirmed three planets in the system were recently used to infer the orbital inclinations and hence the  masses from RV measurements spanning over about 7500\,d \citep{2024arXiv240711520V}. 

Additionally, from the point of view of planetary system formation and evolution, planets in multiple systems are very interesting, because comparative planetology allows constraints for formation and evolution models. An example is the TRAPPIST-1 system, where precise mass and radius determinations of all planets indicate the formation out of a slightly metal poor proto-planetary disk \citep{2021PSJ.....2....1A}. Other examples include systems such as HD\,15337, which contains two planets \citep{Gandolfi_2019} with different composition on both sides of the radius gap \citep{2018ApJ...867...75D}, and both GJ\,876 and TOI\,216, which host
planetary systems that exhibit mean motion resonances \citep{2018ApJ...867...75D,2022ApJ...925...38N}. A total of 326 multi-planet systems with at least three planets are currently known (\href{https://exoplanetarchive.ipac.caltech.edu/cgi-bin/TblView/nph-tblView?app=ExoTbls&config=PS}{NASA Exoplanet Data from August 27, 2024}). For those which do not transit, modelling of the planet-planet interaction using RV data can reveal their  masses, given a sufficiently long observational time base. 

\begin{table}[ht]
   \caption{\label{tab:stellar-params}Stellar parameters for GJ\,1061.}
   \centering
   \begin{tabular}{lcc}
      \hline
      \hline 
      \noalign{\smallskip}
      Parameter & Value & Ref. \\
      \hline 
      \noalign{\smallskip}
      $\alpha$ & 03 35 59.700 & {\em Gaia}\\
      $\delta$  & $-$44 30 45.731 & {\em Gaia}\\
      $\mu_\alpha \cos{\delta}$ [mas/yr]& $745.654\pm 0.035$ & {\em Gaia}\\
      $\mu_\delta$ [mas/yr]& $ $-$373.233 \pm 0.038$ & {\em Gaia}\\
      $\pi$ [mas] & $272.1615 \pm 0.0316$ & {\em Gaia}\\
      $V$ [mag] & $13.06 \pm 0.07 $ & G14\\
      $J$ [mag] & $7.523 \pm 0.02 $ & 2MASS \\
      \noalign{\smallskip}
      Sp. type & M5.5V & H02\\
      $T_{\rm eff}$ [K] & $2953\pm98$, $2999\pm41$ & S18\\
      $L$ [10$^{-3}L_{\odot}$] & $1.7 \pm 0.1$, $3 $ & S18\\
      $R$ [$R_{\odot}$] & $0.156 \pm 0.005$,  $0.19$ & S18\\
      $M$ [$M_{\odot}$] & $0.12 \pm 0.01$, $0.14 $ & S18\\
      {[Fe/H]} [dex] & $-0.08 \pm 0.08$ & N14\\
      \noalign{\smallskip}
      $\gamma$ [km/s] & 1.49 $\pm$0.23 &  {\em Gaia}\\
      $v\sin i$ [km/s] & $<2.5$ & D20 \\
      $\log L_{\mathrm{H}\alpha}/L_\mathrm{bol}^{\rm *}$ & $<$-4.88 (inactive) & D20\\
      $\log L_\mathrm{X}/10^{-7}$\,W &  26.07  & S04 \\
      Age [Gyr]  &    $>7.0\pm0.5$ & W08\\
      $P_{\rm rot} [d]$ & $\gtrsim 125$ & this work\\
      \noalign{\smallskip}
      \hline 
   \end{tabular}
\tablefoot{$^{\rm *}$: see \citet{2018A&A...614A..76J} for a conversion from equivalent width to the luminosity ratio as well as for the threshold for inactive stars.
    }
    \tablebib{
      {\em Gaia}{}: \citet{2023A&A...674A...1G};
      G14: \cite{Gaidos2014MNRAS.443.2561G};
      2MASS: \citet{Skrutskie2006AJ....131.1163S};
      H02: \citet{Henry2002AJ....123.2002H};
      S18: \citet{2018AJ....156..102S};
      N14: \cite{Neves2014A&A...568A.121N};
      D20: \cite{2020MNRAS.493..536D};
      S04: \cite{Schmitt2004A&A...417..651S};
      W08: \cite{West2008AJ....135..785W}.
   }
\end{table}

Finally, the planets orbiting stars closest to our Solar System are the optimal targets for follow-up study of atmospheric chemistry.  The mid M-dwarf GJ\,1061 with a mass of 0.12\,M$_\odot$ (Table\,\ref{tab:stellar-params}), is of particular interest: (i) The system harbours three potentially rocky planets, two of them in the conservative habitable zone, detected using the RV technique: GJ\,1061\,b: P=$3.204\pm 0.001$\,d, $M\sin i = 1.37\pm0.16 M_\odot$, GJ\,1061\,c: P=$6.689\pm0.005$\,d, $M\sin i = 1.74\pm0.23 M_\odot$, GJ\,1061\,d: P=$13.03\pm0.03$\,d, $M\sin i = 1.64\pm0.24 M_\odot$  \citep{2020MNRAS.493..536D}, (ii) it is located at a distance of only 3.6\,pc making it the closest compact multi-planetary system to our Solar System with at least three planets.
The radii of the planets are unknown, as the planets do not transit \citep{2022A&A...665A.157L}.  The planetary system is dynamically packed, so that planet-planet interactions may be detectable using intensive RV monitoring, and we have sufficient data taken over a long baseline to provide a robust solution. 

In this paper, we investigate the gravitational planet-planet interactions of the three planets orbiting the mid-M dwarf GJ\,1061.  In Section\,\ref{sect:Observations} we describe the observations and data processing, and in Section\,\ref{sect:Methods} we specify the methods and computer codes used in this work. In Section\,\ref{sect:Results} we present our results and interpret them.

\section{Observations and data}
\label{sect:Observations}

\subsection{Spectroscopic data}
\label{sect:spec_data}
For the measurement of the RV variation, we use data that is publicly available on the ESO Science Archive\footnote{\url{https://archive.eso.org/wdb/wdb/adp/phase3_spectral/form}} from the HARPS\footnote{High Accuracy Radial velocity Planet Searcher} instrument \citep{2003Msngr.114...20M} at the 3.6m telescope of La Silla observatory, Chile. Table \ref{tab:obs_campaigns} lists the five campaigns that comprise the 184 observations of the full dataset. A total of seven spectra were recorded prior to the fibre change of the instrument in June 2015 and 177 post fibre change but before the 2020 warm-up. We treat these subsets of HARPS data as distinct datasets in this analysis.\par

The spectroscopic observations were reduced for both subsets separately using the \texttt{serval} algorithm, \citep{2018A&A...609A..12Z} which uses an iterative template matching process to first obtain RVs relative to the template and then recalculate the template by co-adding the RV corrected spectra. It begins with the highest signal-to-noise (S/N) spectrum as the template, and is programmed to stop after the first co-added template is created and matched.  Three spectra were discarded during this process by \texttt{serval} for their low S/N, leaving 7 pre- and 174 post-fibre change spectra. By default, \texttt{serval} further discards the first ten HARPS echelle orders due to their lower S/N. Due to the late spectral type of GJ\,1061 and correspondingly low flux in the blue-most orders, we decided to restrict this further to orders 20 to 72.\par
Outliers were removed by applying a conservative 5-$\sigma$ clipping, leaving 172 individual RV measurements. Two RV datasets were created: (1) averaged in the case of multiple observations per night, resulting in a final total of 142 data points, and (2) an unbinned dataset sensitive to any ultra-short period signals.

The HARPS data were complemented with 26 spectra from ESPRESSO\footnote{Echelle SPectrograph for Rocky Exoplanets and Stable Spectroscopic Observations} \citep{2021A&A...645A..96P}. We also used \texttt{serval} to retrieve the RV measurements, again skipping the bluest orders and using orders 16 up to 169. A 5-$\sigma$ clipping had no effect, the nightly averaging reduced the number of measurements to 25.

Along with the RV measurements, {\tt serval} determines spectroscopic activity indicators including the H$\alpha$ equivalent width, the chromatic index, and the differential line width (dLW). More details are described in \citet{2018A&A...609A..12Z}. 

\begin{table}
\centering
\caption{Observing programs that contributed to this work. The spectra of the first two were taken with HARPS prior to the fibre upgrade, the next three after the upgrade, and the last one with ESPRESSO.}
\label{tab:obs_campaigns}
\begin{tabular}{ccccl}
\hline\hline
ESO ID & PI & Nr. Spectra & median error\\
       &    &             & m/s\\
\hline
072.C-0488(E) & Mayor, M. & 4 & \multirow{2}{*}{1.66} \\
183.C-0437(A) & Bonfils, X. & 3 & \\[0.3cm]
0101.C-0516(A) & Jeffers, S. & 56 & \multirow{3}{*}{1.41}\\
1102.C-0339(A) & Bonfils, X. & 56 & \\
198.C-0838(A) & Bonfils, X. & 65 & \\[0.3cm]
0112.C-2112(A) & Jeffers, S. & 26 & 0.25\\
\hline\\
\end{tabular}
\end{table}

\subsection{Photometric data}
\label{sec:phot_data}
GJ\,1061 has been observed with MEarth\footnote{\url{https://lweb.cfa.harvard.edu/MEarth/Welcome.html}}, a photometric monitoring project to search for transiting planets around $\sim$3000 nearby mid-to-late M dwarfs, each using eight 40\,cm-telescopes in the northern and southern hemisphere \citep{2012AJ....144..145B}. GJ\,1061 was observed by Telescopes 13 and 11 of MEarth-South at Cerro Tololo Inter-American Observatory (CTIO) in Chile from June 1, 2017 to  February 27, 2022 and from October 20, 2016 to March 10, 2020, respectively. The 18948 and 53510 epochs were filtered with a 5-$\sigma$ clipping and then binned to 663 and 276 nightly averages. 

There are also photometric observations from the  All-Sky Automated Survey for Supernovae \citep[ASAS-SN;][]{Shappee2014, Kochanek2017, 2023MNRAS.519.5271C}. We retrieved the light curve from their online server \footnote{\url{https://asas-sn.osu.edu/}}, which does not include proper-motion correction. We therefore corrected the coordinates for each year of observation (from 2014 to 2024), which were specified in each query, and then downloaded the data and combined them into a single (corrected) light curve. The observations were made with the $V$ and $g$ filters using different cameras.  

As with MEarth, we also binned the ASAS-SN data nightly ($1$\,d) and performed a 5-$\sigma$ clipping. After this, the cameras bE, bm, and bi (all $g$ filters) had only one data point, so we discarded them. The bn camera ($V$ filter) was also discarded as it had only 44 points with a time span of $50$\,d, covering the same time series as the bf camera (also $V$ filter). Both the MEarth and ASAS-SN time series are shown in Fig.\,\ref{fig:phot_timeseries}. Note that the MEarth-11 data follows a long-term trend, and that the MEarth-13 dataset has a long observation gap ($\sim 700$\,d). 

\begin{figure}
    \centering
    \includegraphics[width=0.49\textwidth]{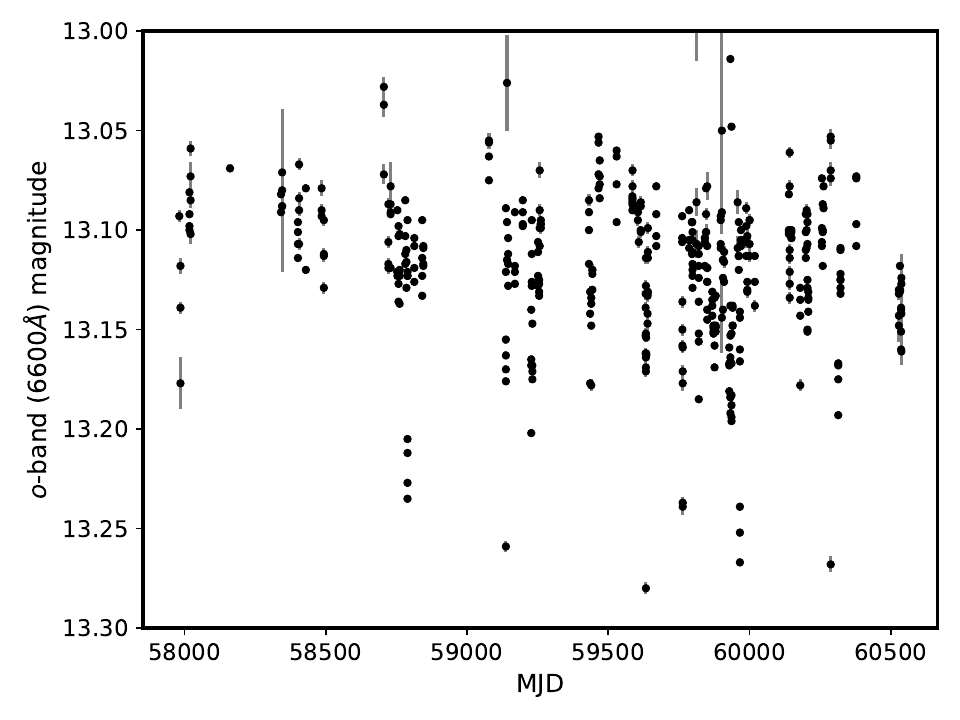}
    \caption{ATLAS $c$-band light curve of GJ~1061.  Some error bars are smaller than the points.}
    \label{fig:atlas-lc}
\end{figure}

\begin{figure}
    \centering
    \includegraphics[width=0.495\textwidth]{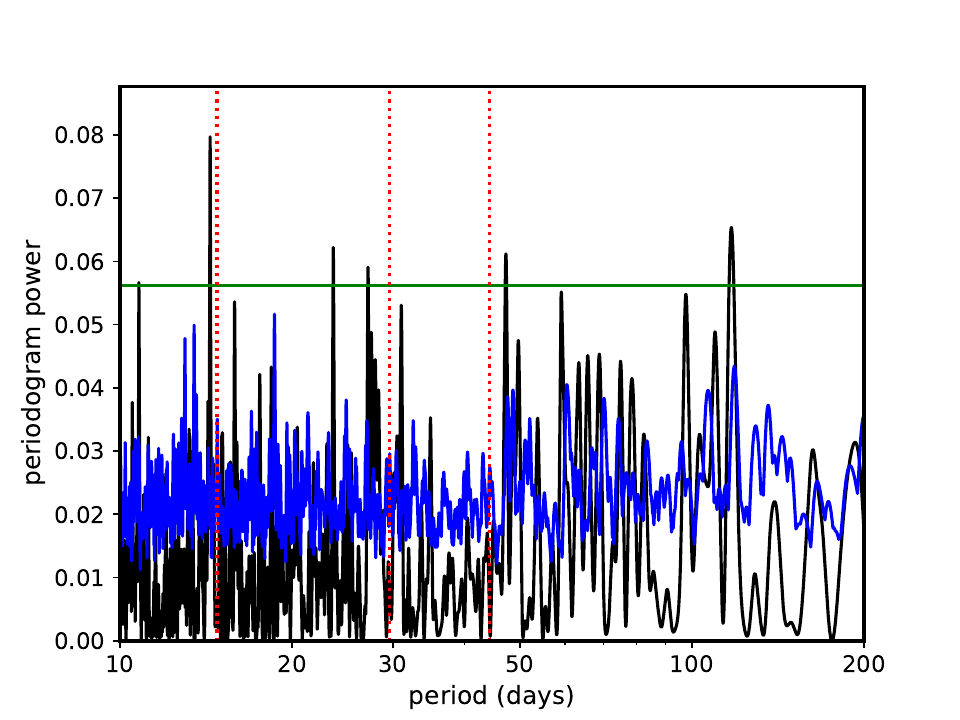}
    \caption{Lomb-Scargle periodogram of the ATLAS $c$-band photometry of GJ~1061.  The green-line is the $p=0.01$ false-alarm threshold calculated using, and the blue curve is the maximum of the ensemble of 100 periograms of the randomly scrambled photometry.  The red dotted lines mark the lunar synoptic period and its 1:2 and 3:2 harmonics.}
    \label{fig:atlas-pg}
\end{figure}

We also retrieved "cyan" ($c$-band, 5200\aa) and "orange" ($o$-band, 6600\aa) forced photometry from the Asteroid Terrestrial-impact Last Alert System \citep[ATLAS,][]{Tonry2018,Heinze2018}.  Due to the high proper motion (0.83" yr$^{-1}$) of GJ\,1061 compared to the 1.86" pixel scale of ATLAS, we performed the retrieval in 15 six-month intervals. GJ\,1061 is saturated (median magnitude of 11.2) in $o$-band but marginally not in $c$-band (13.1) and we performed a Lomb-Scargle periodogram analysis \citep{Scargle1982} on the 444 points in the $c$-band light curve with a median precision of 0.002 mags (Fig. \ref{fig:atlas-lc}).  The periodogram (Fig. \ref{fig:atlas-pg}) contains a marginally significant peak at a $\sim$125\,d signal, which might be the rotational signal tentatively observed in the RV analysis.

An overview of the photometric data is listed in Table\,\ref{tab:phot-data}.

\section{Methods}
\label{sect:Methods}
The measured RV values were analysed using a multi-planetary model, employing either Keplerian or Newtonian orbits. In the Keplerian case, the radial velocity is derived analytically, while in the Newtonian case, it is obtained through direct N-body integration using {\tt rebound} with the {\tt IAS15} integrator \citep{2012A&A...537A.128R,2015MNRAS.446.1424R}. The key difference is that in Keplerian models, each planet's orbit is calculated as though it were the only planet in the system, whereas Newtonian models explicitly account for the mutual gravitational interactions between all bodies. Small perturbations due to these interactions manifest as apsidal precession. Over time, this causes a Keplerian orbit to exhibit a phase shift relative to the observed RV orbit. The strength of the perturbation depends on the planetary masses, while the RV amplitude is determined by the planetary mass and orbital inclination. Due to the perturbations, the orbital elements are not constant but evolve with time. Energy and angular momentum exchange lead to variations in the orbital period and eccentricity. Therefore, RV measurements over a sufficiently long time baseline allow for the determination of planetary masses and inclinations, constrained by both the RV amplitudes and the phase drift of the orbital motions.

The Keplerian model is parameterised with the semi-amplitude $K$, the orbital period $P$, the eccentricity $e$, the longitude of periastron $\omega$, and the mean longitude $\lambda$. While the inclination $i$ and the longitude of the ascending node $\Omega$ remain undetermined in the Keplerian model, both can in principle be determined in the case of dynamical orbits from the strength of the planet-planet gravitational interaction. For Newtonian orbits, we replace $K$ by the planetary mass $m$ and add $i$ and $\Omega$ as additional free parameters. It should be noted that the dynamical model can only provide $\Omega$ relative to e.g. that of the innermost planet, where $\Omega=0$ is assumed. In the scenario where planet-planet interactions are weak, a simplification assuming co-planar orbits might be appropriate. In that case, only a common inclination is determined and $\Omega=0$ is set for all planets.

As discussed in Liebing et al. (A\&A, accepted), the parametrisation of $e$ and $\omega$ has an impact on the posteriors of other planetary parameters. We therefore used either $e$ and $\omega$ directly or alternatively $h^{\prime}=\sqrt e\sin\omega$ and $k^{\prime}=\sqrt e\cos\omega$ to check the impact of the parametrisation on the results. Since the planetary system of GJ\,1061 is compact, we expect small eccentricities in order to keep it dynamically stable. 

For the Newtonian orbits, we check the critical two-planet separation in units of the mutual Hill radius \citep[Equation 5,][]{2017Icar..293...52O}. We reject parameter combinations violating this critical Hill-radius separation criterion to avoid computational time being spent on physically unlikely parameters. 

The planetary model of varying complexity is complemented with Gaussian Process regression (Sect.\,\ref{sect:GP}) to model possible rotational modulated contributions from stellar activity. The posterior distributions of the parameters and hyperparameters are determined with a Nested Sampling algorithm (Sect.\,\ref{sect:NS}) which also provides the Bayesian evidence which is used to identify the best model (Sect.\,\ref{sect:bestmodel}). Finally, the parameters of the best model are checked for orbital stability (Sect.\,\ref{sect:SPOCK})

\subsection{Gaussian process regression}
\label{sect:GP}
Gaussian Process (GP) regression is a powerful non-parametric Bayesian method for modelling complex functions. GPs provide a flexible way to model data by assuming that the underlying function is a sample from a Gaussian distribution. The python package \citep[\texttt{celerite,}][]{2017AJ....154..220F} is an efficient implementation of GP regression that is specifically designed for large datasets, since its computational demands scale linearly with the number of data points. This requires that the covariance matrix (the kernel) is modelled as a sum of complex exponential functions. \citet{2017AJ....154..220F} suggest using a simple damped harmonic oscillator ({\em SHO}) as kernel in case of modelling the covariance caused by stellar activity. Mathematical details are given in Equations 9, 20, and 21 in \citet{2017AJ....154..220F}. An improved kernel for rotationally modulated stellar activity is presented by \citet{celerite2} and it consists of two {\em SHO} kernels with the period of the second being the second harmonic of the first. This {\em dSHO} kernel is parameterised by the variance $\sigma$, the oscillator (rotation) period $P$, the quality factor $Q_0$ and the difference of the quality factor of the first and second period $dQ$, which both represent the degree of damping, and the fractional amplitude between the two oscillators $f$.
The resulting damping timescale is enforced to be the same for both oscillators. We use wide priors (see Table\,\ref{tab:GP}) for these parameters and require that $Q>1/2$ to keep the oscillators in the under-damped regime. We also allow for additional white noise using a jitter term, i.e. a covariance matrix with diagonal elements only. 

\subsection{Nested sampling}
\label{sect:NS}
Nested sampling \citep{2004AIPC..735..395S} is a Bayesian inference method that can be used to estimate the logarithm of the evidence ($\ln \mathcal{Z}$) and the posterior distributions of a model. The python package \texttt{dynesty} \citep{2019S&C....29..891H,2020MNRAS.493.3132S} is an implementation of nested sampling that can be used to perform parameter estimation and model selection. It can be used to efficiently explore complex posterior distributions, and it has a user-friendly interface that makes it easy to use in a variety of applications. One of the key features of {\tt dynesty} is its dynamic nested sampling algorithm, which allows it to adapt to the structure of the posterior distribution in real-time, making it more efficient than other nested sampling methods. Additionally, the package provides a wide range of diagnostic tools and options for controlling the sampling process, enabling users to fine-tune the method to their specific needs. 

We used 5000 live points, \texttt{bound=multi}, \texttt{sample=rwalk}, and stop when $\delta \ln \mathcal{Z}<0.01$. The model selection was made using the \texttt{static} sampler, while the final model was run with the \texttt{dynamic} sampler, optimising the posterior distributions. Multiple runs for model D2, the Keplerian three-planet model with a uniform prior for the period of the {\em SHO} kernel (see Table\,\ref{tab:mod_comp}), resulted in very similar posteriors, indicating that the number of live points were sufficient.

\subsection{Post-processing using \texttt{SPOCK}}
\label{sect:SPOCK}
 The Stability of Planetary Orbital Configurations Klassifier \citep[\texttt{SPOCK},][]{2020PNAS..11718194T} is a machine learning algorithm that is designed to predict the long-term stability of planetary systems over $10^9$ orbital periods of the innermost planet. The algorithm is trained on a set of numerical simulations of about 100\,000 nearly co-planar planetary systems with three or more planets with masses ranging from Mars- to twice Neptune-mass. It uses a combination of physical and dynamical features to classify the stability of a given system. It has been used to predict the stability of observed exoplanetary systems \citep[e.g.][]{2020PNAS..11718194T,2022MNRAS.510.5464K}, and it has been shown to be accurate in identifying stable systems. Additionally, it has been used to identify new configurations of exoplanetary systems that are likely to be stable, which can help guide the search for new exoplanets.
 
 We used {\tt SPOCK} to check the resulting planetary system configurations for dynamical stability. We emphasise that we use the posterior distributions and not draws from the normal distributions given by the mean and standard deviation tabulated in Table\,\ref{tab:mod_dyn}.

\subsection{Light curve modelling}
\label{sect:lightkurve}
To obtain the stellar rotation period we performed a GP fit on all the data with \texttt{juliet} \citep{Espinoza2019}, which provides Keplerian components by fitting RVs \citep[\texttt{Radvel},][]{Fulton2018}, transits for photometric models \citep[\texttt{batman},][]{Kreidberg2015}, and a red noise component (GP) using the publicly available packages \texttt{celerite} and \texttt{george} \citep{Ambikasaran2015}. We use the {\em dSHO} kernel for the modelling of the stellar rotation modulation, fitting the GP hyperparameters together with a linear trend for each dataset. 

\section{Analysis results}
\label{sect:Results}

\subsection{Stellar rotation period}
\label{sect:rotation}
We computed the Generalized Lomb-Scargle (GLS) periodogram \citep{2009A&A...496..577Z} of the activity indicators and the photometric data (Fig.\,\ref{fig:phot_timeseries}).  In  Fig.\,\ref{fig:GLS_all_GJ1061_paper} we display the periodograms which have signals reaching at least a  FAP level of 10\% (excluding yearly and half-yearly periodicities). In addition to the signal at $1$\,d caused by nightly observations, the most dominant signal (FAP $<$ 0.1\%) corresponds to $130$\,d in one of the ASAS-SN datasets (bf cam, $V$ filter), which is also present in the MEarth-13 data, but the latter has several significant periodicities between $100$\,d and $200$\,d. Other significant signals (FAP $<$ 0.1\%) are found in the MEarth-11 data, at $160$\,d, which become stronger after fitting a trend (see Section 2.2). At $200$\,d there are also peaks in the ASAS-SN bn and bF cam ($g$ filter), as well as in MEarth-11, the latter also showing a dominant peak around $300$\,d. These long periodicities could be related to a long magnetic cycle, since stellar rotation periods have shorter timescales \citep{2024A&A...684A...9S}.

 We used the dynamic nested sampling algorithm from \texttt{dynesty} (see Section\,\ref{sect:NS} and Fig.\,\ref{fig:GLS_all_GJ1061_paper}) and chose a dSHO kernel with uniform priors on the period from $1$\,d to $200$\,d (the rest of the priors are listed in Table\,\ref{tab:priors_Prot}).  We first fitted a GP to each photometric dataset separately, obtaining only a clear detection of the GP period on MEarth-11 ($\sim160$\,d, trend fitted), MEarth-13 ($\sim46$\,d) and ASAS-SN bf cam ($V$ filter, $\sim137$\,d). 
Multiple star spots at different longitudes can result in sub-rotational period signals, which could be the reason for the detection of shorter periods in some datsets. A possible activity cycle like in our Sun could lead to non-detections of photometric variability in some datasets. In this sense, we cannot retrieve a reliable stellar rotation period, but we can make an estimate. Firstly, we do not rely entirely on the MEarth-13 data alone because of the long gap. Since MEarth-11 and ASAS-SN bf cam ($V$ filter), which are the only remaining datasets where a periodicity has been found, give long periods ($P > 120$\,d), we expect a rotation period of this order. In this sense, we can rely on the combination of the MEarth-11 (trend fitted) with the ASAS-SN bf cam $V$ filter as an estimate of the stellar the rotation period. We then combined MEarth-11 (trend fitted) with ASAS-SN bf cam ($V$ filter) and obtained a stellar rotation period of $130^{+5}_{-14}$\, d. 

\begin{figure}
    \centering
    \includegraphics[width=0.49\textwidth]{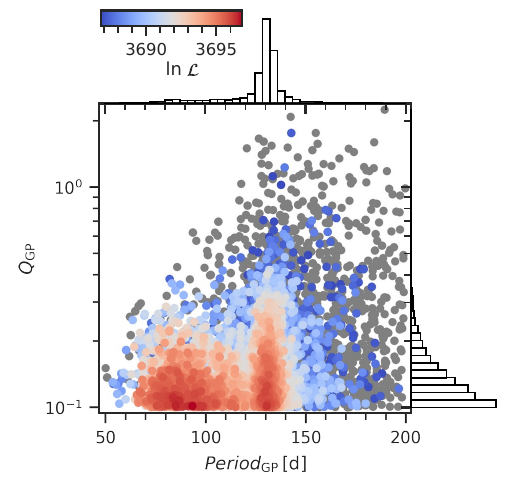}
    \includegraphics[width=0.49\textwidth]{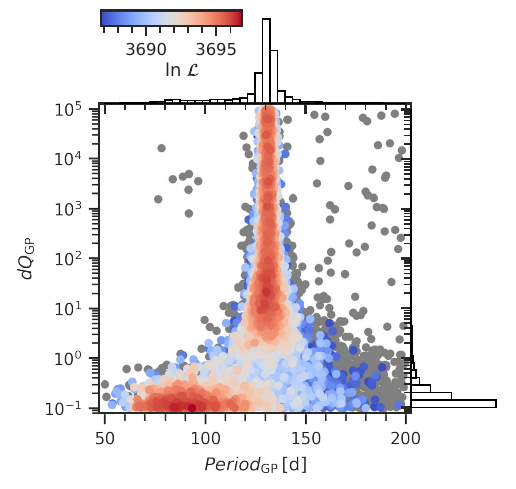}
    \caption{Hyperparameters of the GP model for the photometric data: The quality factor $Q$ versus the rotation period (top) and the difference in quality factor between primary and secondary oscillator (bottom). The colour indicates the likelihood.}
    \label{fig:GP-vs_period_MEarth-11_bf}
\end{figure}

The posteriors are listed in Table\,\ref{tab:posteriors_Prot}, and we also depict the GP hyperparameters ($dQ$ and $Q$) of the dSHO kernel against the periodic component in Fig.\,\ref{fig:GP-vs_period_MEarth-11_bf}. The likelihood and number of posterior samples favour a rotation period of $\sim 125$\,d. A better sampled photometric dataset (i.e. with no block separation, sampled for at least three probable rotation periods) is required in order to have an unambiguous detection. 

The spectroscopic activity indicators, namely the H$\alpha$ equivalent width probing the chromospheric activity, the chromatic index indicating RV variations do to changing spot contrast, and the dLW characterizing line shape variations (\ref{sect:spec_data}), do not show a significant signal near the potential rotation period or its harmonics. Only the latter shows a weak signal at about 125\,d (Fig.\ref{fig:GLS_all_GJ1061_paper}).

The inferred rotation period of $\sim 125$\,d is compatible with the low rotational velocity $v \sin i < 2$\,km\,s$^{-1}$ \citep{2020MNRAS.493..536D} and the estimate age of older than 7.0\,Gyr \citep{2008AJ....135..785W} and in agreement to the estimated rotation period from the previous analysis \citep{2020MNRAS.493..536D}. The As discussed below, the RV data also show a signal at 125\,d, which we interpret as the rotational induced RV red noise. The corresponding modulation is shown in Fig.\,\ref{fig:RVplot} as a thin grey line. Its amplitude is below 2\,m\,s$^{-1}$

\begin{figure}
    \centering
    \includegraphics[width=0.499\textwidth]{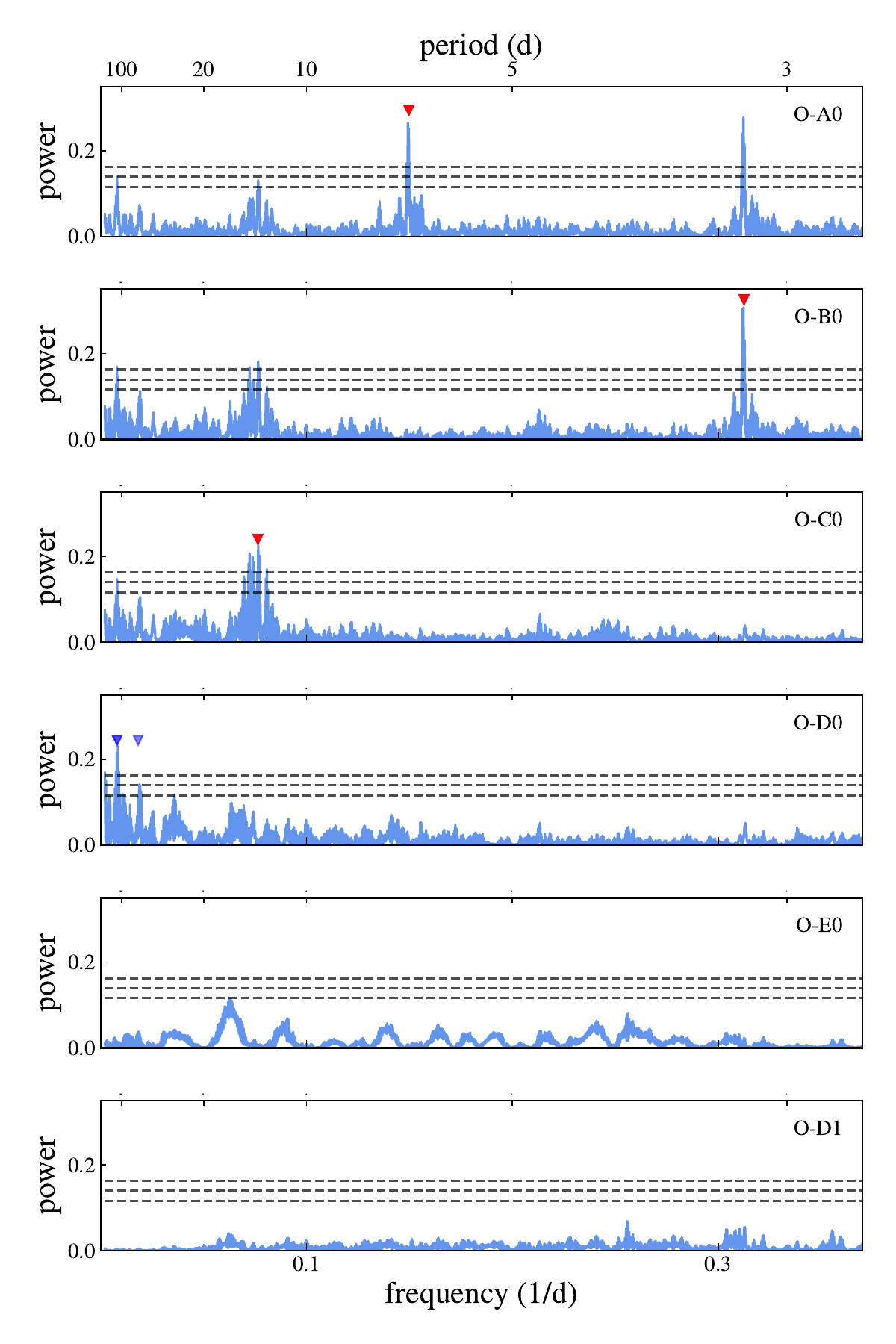}
    \caption{Periodograms of models with increasing complexity: From top to bottom: 0, 1, 2, 3, planets, and 3 planets + {\em SHO} kernel. Red arrows mark the subtracted signal for the following periodogram, the blue arrow indicates the potential stellar rotation period at $\sim$125\,d, the light blue arrow points at the 55\,d signal. The labels indicate the model which is subtracted from the observation "O" (see Table\,\ref{tab:mod_comp}).}
    \label{fig:GLSperiodogram}
\end{figure}
\begin{figure}
    \centering
    \includegraphics[width=0.53\textwidth]{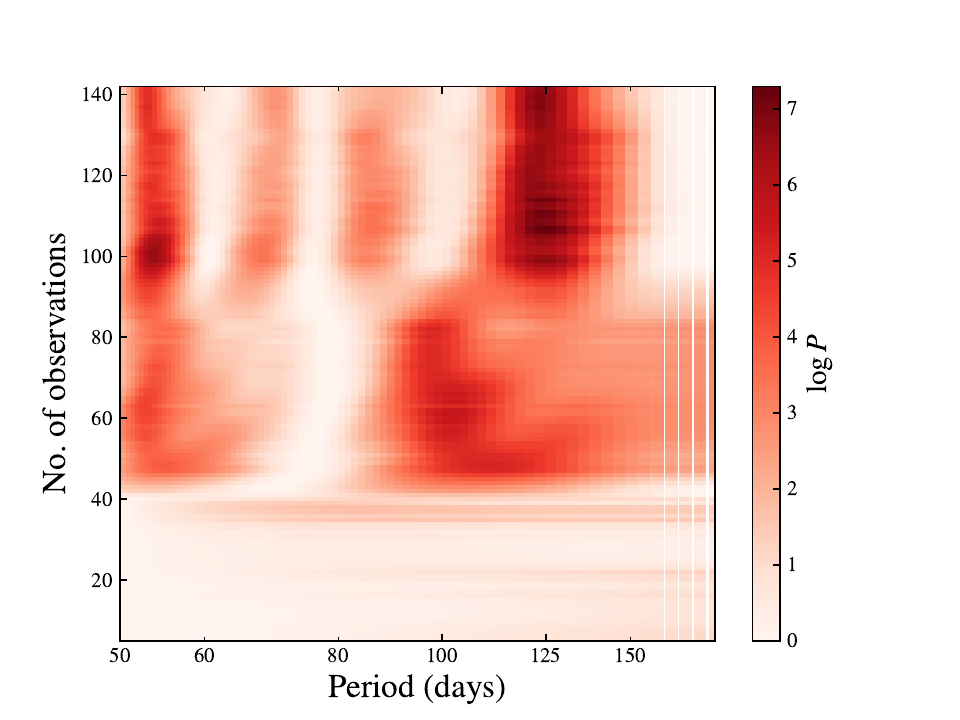}
    \caption{Stacked Bayesian Periodogram (s-BGLS) of the HARPS post-fibre upgrade data after subtraction of 3 planets (model D0).}
    \label{fig:sbgls}
\end{figure}

\begin{table}
    \centering
    \caption{Model comparison for model with increasing number of potential planets. }
    \begin{tabular}{c@{}c@{}c@{}c@{}c@{}c@{}c@{}c@{}c@{}c@{}}
    \hline
        &\multicolumn{5}{c}{Planets} & GP & $\ln \mathcal{Z}$& $\Delta \ln \mathcal{Z}$\\
        Model\,\, &3.2d\,\,\, & 6.7d\,\,\, & 13.1d\,\,\, & 55d\,\,\,& 125d\,\, & \\
        \hline
        A0 &    &      &      &    &     & no        &\,-463.3\,\, &-97.1\\
        B0 &    &  X   &      &    &     & no        &\, -442.4\,\, &-76.2\\
        C0 &X   &  X   &      &    &     & no        &\, -416.8\,\, &-50.6\\
        D0 &X   &  X   &  X   &    &     & no        &\, -397.3\,\, &-31.1\\
        E0 &X   &  X   &  X   &    &  X  & no        &\, -375.3\,\, &-9.1\\
        F0 &X   &  X   &  X   & X  &  X  & no        &\, -371.3\,\, &-5.1\\
        D1 &X   &  X   &  X   &    &     & {\em SHO}\,$\mathcal{N}$ &\, -376.1\,\, &-9.9\\
        D2 &X   &  X   &  X   &    &     & {\em SHO}\,$\mathcal{U}$ &\, -374.6\,\, &-8.4\\
        D3 &X   &  X   &  X   &    &     & {\em dSHO}\,$\mathcal{U}$&\,-377.6\,\, &-11.4\\
        E1 &X   &  X   &  X   &    &  X  & {\em SHO}\,$\mathcal{N}$&\, -374.2\,\, &-8.0\\
        E2 &X   &  X   &  X   &    &  X  & {\em SHO}\,$\mathcal{U}$ &\, -369.8\,\, &-3.6\\
        E3 &X   &  X   &  X   &    &  X  & {\em dSHO}\,$\mathcal{U}$&\, -372.7\,\, &-6.5\\
        F1 &X   &  X   &  X   &  X &  X  & {\em SHO}\,$\mathcal{N}$ &\, -371.6\,\, &-5.4\\
        F2 &X   &  X   &  X   &  X &  X  & {\em SHO}\,$\mathcal{U}$ &\, -371.4\,\, &-5.2\\
        F3 &X   &  X   &  X   &  X &  X  & {\em dSHO}\,$\mathcal{U}$&\, -371.2\,\,&-5.0\\
         \hline 
        D4 &X   &  X   &  X   &    &     & {\em SHO}\,$\mathcal{N}$ &\, -366.1\,\, & 0.1 \\
        D5 &X   &  X   &  X   &    &     & {\em SHO}\,$\mathcal{U}$ &\, -366.2\,\, & 0 \\
        D6 &X   &  X   &  X   &    &     & {\em dSHO}\,$\mathcal{U}$ &\, -371.8\,\, & -5.6 \\
        E4 &X   &  X   &  X   &    &  X  & {\em SHO}\,$\mathcal{U}$ &\, -376.9 \,\,&-10.7\\
        F4 &X   &  X   &  X   &  X &  X  & {\em SHO}\,$\mathcal{U}$ &\, -376.6\,\, & -10.4\\
         \hline          
         \end{tabular}
    \tablefoot{The planets included are marked with "X" under the approximate period of the signal. The first block corresponds to models using Keplerian orbits, the second to Newtonian models. $\mathcal{N}$ and $\mathcal{U}$ denote the priors for the GPs: Normal prior $\mathcal{N}(125,5)$ around the potential rotation, and uniform prior $\mathcal{U}(30,200)$.}
    \label{tab:mod_comp}
\end{table}

\subsection{Planetary model selection}
\label{sect:bestmodel}
In a first analysis step, we selected the model with the highest $\log \mathcal{Z}$. As a selection criterion, we used the difference in the natural logarithm of the Bayesian evidence $\Delta \ln\mathcal{Z}$. Following \citet{2008ConPh..49...71T}, we use the threshold of $\Delta \ln \mathcal{Z} > 5$ as strong evidence and $\Delta \ln \mathcal{Z} > 2.5$ as moderate evidence for a more complex model being better than a simpler model. For the model selection we used Keplerian models since the deviation between Keplerian and Newtonian models are rather small so that we do not expect significant differences in the model selection caused by this simplification. 

Starting with our base model which allows for offsets and jitter terms for the two HARPS and the ESPRESSO datasets separately (Model A0, Table\,\ref{tab:mod_comp}), we recursively add a Keplerian orbit at the frequency with the highest power in the GLS periodogram \citep{2009A&A...496..577Z} of RV residuals until $\mathcal{Z}$ is no longer increasing (Models B0 to F0, Table\,\ref{tab:mod_comp}) and no peak in the periodogram is above the 1\% false alarm probability. We use narrow priors for the three known planets with a width of the frequency resolution given the length of the dataset. After adding five Keplerian signals to the model, no significant power is left in the GLS periodogram of the residuals. The first three of these Keplerian models are the already known planets at 6.68\,d, 3.20\,d and 13.1\,d (Models B0 to D0) from \cite{2020MNRAS.493..536D}. The signal at 55\,d has been identified as a planet candidate by \citet{2020MNRAS.493..536D}. In Fig.\,\ref{fig:GLSperiodogram} we show periodograms where we have successively subtracted the planet signature and the red-noise modulation.

Instead of interpreting the 55\,d signal as a planet, the long period signals were modelled with a {\em SHO} and {\em dSHO} kernel. This is motivated by the varying period and power of the Stacked Bayesian GLS (s-BGLS) periodogram \citep{2017A&A...601A.110M} of signals in the long-period range above 50\,d (Fig.\,\ref{fig:sbgls}), indicating activity induced radial velocity variations. The 55\,d and the 125\,d periods are also close to the potential rotation period or its second harmonic (Sect.\,\ref{sect:rotation}). In particular, the strength of the 55\,d signal declines after N$_{\rm obs}\sim$100 as more observations are added. The comparison shown in Table\,\ref{tab:mod_comp} reveals that the more complex {\em dSHO} kernel is not better than the {\em SHO} kernel when using wide priors for the rotation period. The resulting periodicity of about 125\,d is close to the rotation period (Table\,\ref{tab:GP}, Fig.\,\ref{fig:corner}) estimated from the analysis of the photometry (Sect.\,\ref{sect:rotation}). A normal prior around the potential stellar rotation period results in a slightly lower $\mathcal{Z}$ compared to a wide uniform prior. Models with Gaussian process regression are significantly better ($\Delta \ln \mathcal{Z}>5$) compared to those without for three and four planets (Models D0 versus D2 and E0 versus E2). In the case of the five planet models (F0 versus F2), all have nearly identical evidence. Together, this indicates that we can confirm the three known planets of GJ\,1061 \citep{2020MNRAS.493..536D}. The additional signals at 55\,d and 125\,d are more likely due to stellar rotation, although the five-planet model (F0) is statistically moderately better than the three-planet model including the {\em SHO} kernel (D2) to model the rotational RV variability. No additional planets can be detected in the current data.

Having established the best Keplerian model, we now model the planet-planet interaction. We do this by using Newtonian models (see Section\,\ref{sect:Methods}) for models with three to five planets using a stellar mass of 0.12\,M$_\odot$, shown in the lower block in Table\,\ref{tab:mod_comp}. We first checked that the parametrisation of the eccentricity and the longitude of periastron has only a minor impact on the derived model parameters. The final model is therefore run with $h^{\prime}=\sqrt(e)\sin\omega$ and $k^{\prime}=\sqrt(e)\cos\omega$ using uniform priors within $[-0.4,0.4]$, the mean longitude $\lambda$ was sampled uniformly within $[0,2\,\pi]$ employing periodic boundaries (see also Table\,\ref{tab:mod_dyn}). We then compared the results from a co-planar (Model D4, D5, D6) to a full three-dimensional (3D) model, where a common inclination and a fixed longitude of the ascending node of $\Omega=0$ for all planets were fitted in the former.  We also used three different GP kernels, the SHO kernel with uniform and normal priors, as well as the dSHO kernel with uniform priors. Like in the case of the Keplerian models, the SHO kernel results in better $\mathcal{Z}$ values. The Newtonian models are always significantly better than the corresponding Keplerian models. The GP kernels do have a slight impact on the posteriors of the planet masses (see Fig.\,\ref{fig:combined_masses}). We therefore use a weighted mean for deriving the mass uncertainties.

The results presented in Table\,\ref{tab:mod_dyn} show that the mutual inclinations derived from the 3D-model are identical within the uncertainties. In combination with the fact that the co-planar model has a significantly higher $\mathcal{Z}$, identifying this as the statistically preferred model, we conclude that deviation from a co-planar configuration are too small to be detected, and we can thus only constrain the inclination of a co-planar configuration. Nevertheless, the planet masses are very similar between the two solutions, indicating that a co-planar model does not introduce major systematic uncertainties. 

To evaluate the impact of the ESPRESSO data, we also used the co-planar model with an identical set-up  of model D5 but restricted it to the HARPS data. The results are listed as a third block in Table\,\ref{tab:mod_dyn}. Especially for the two inner planets, including the ESPRESSO data leads to a decrease in the uncertainty of the mass determination from about 20\% down to 10\% as well as a systematic decrease due to a shift in the inclination.  When the uncertainty of the modeling of the stellar noise is accounted for, the error bars for the planetary masses are slightly larger, especially the upper limits for planets c and d are higher. Nevertheless, their masses are clearly below 2\,$M_\oplus$ and hence likely rocky.

\begin{figure}
    \centering
    \includegraphics[page=3,width=0.495\textwidth]{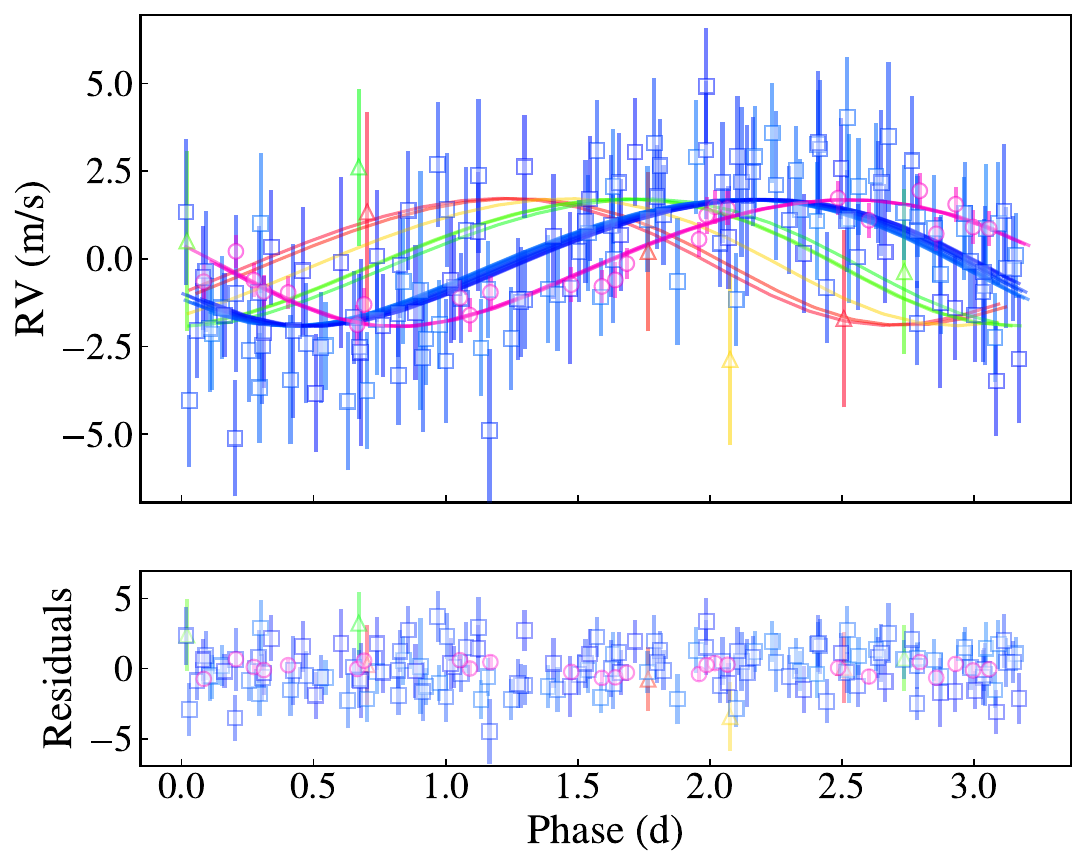}
    \includegraphics[page=5,width=0.495\textwidth]{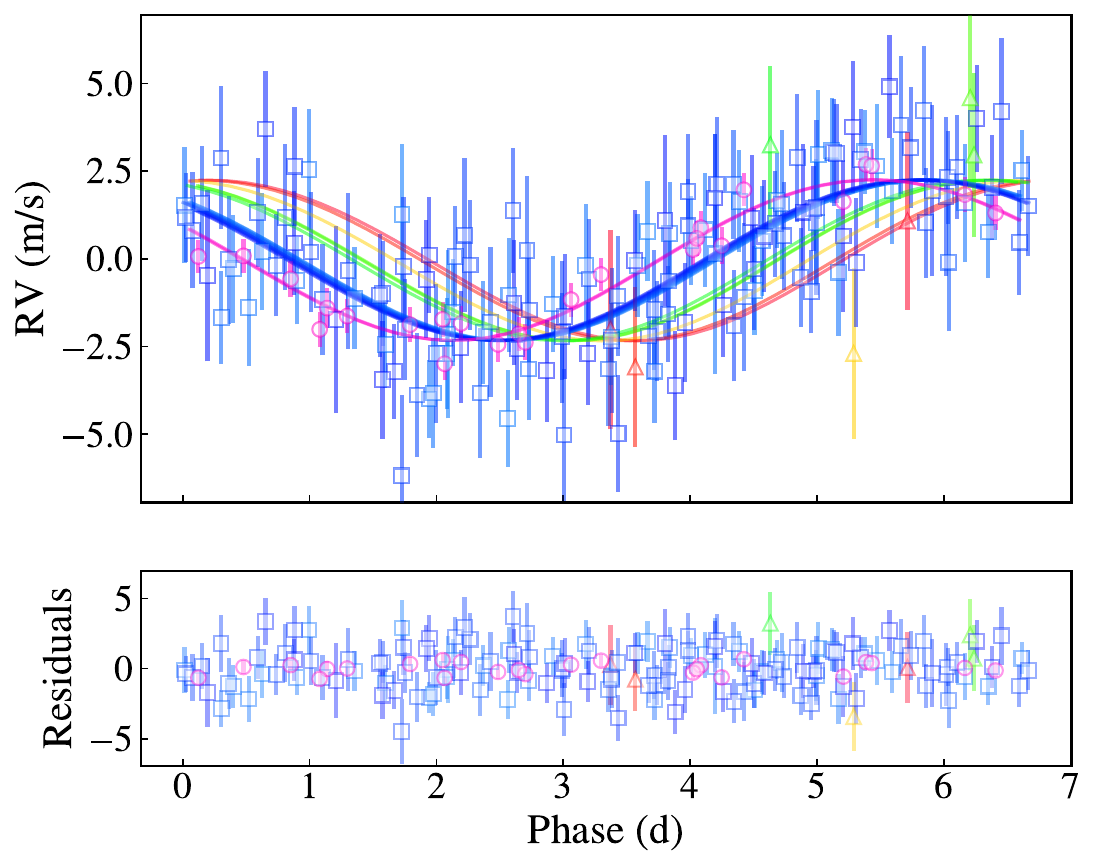}
    \includegraphics[page=7,width=0.495\textwidth]{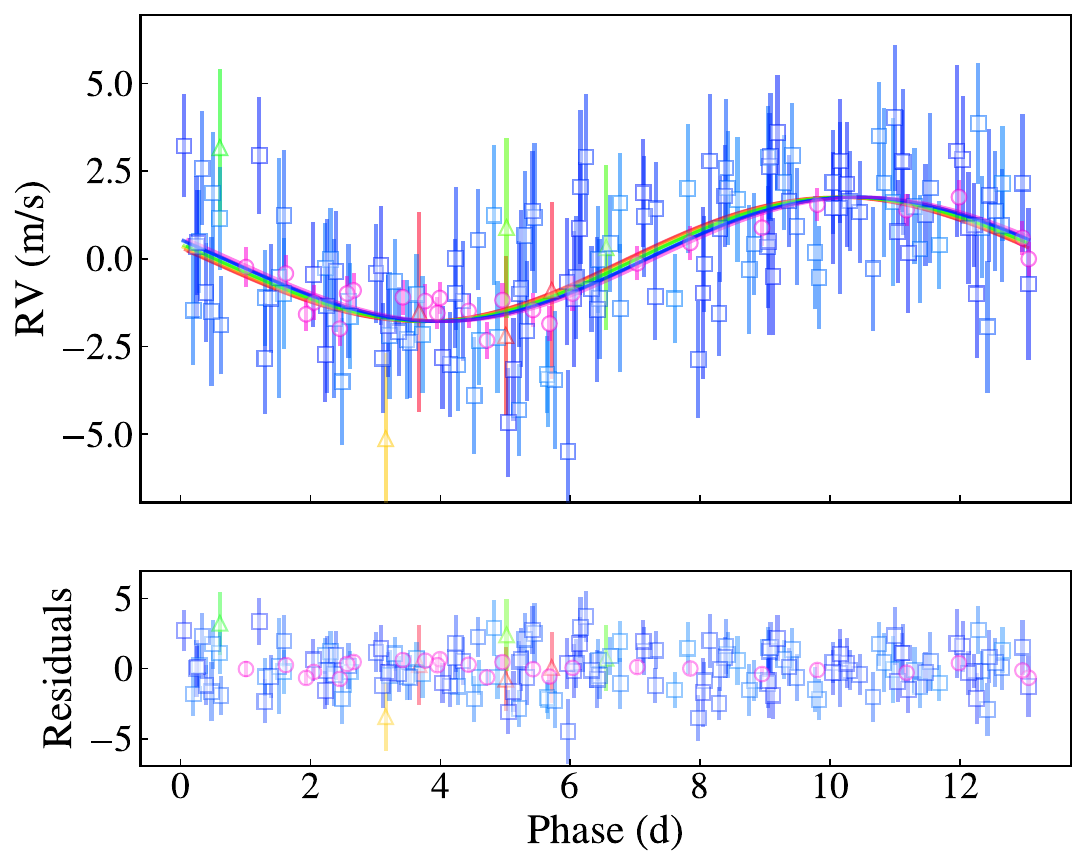}
    \caption{Data versus model D5 (Table\,\ref{tab:mod_comp}) for planets b, c, and d from top to bottom: To show the temporal evolution of the planet's orbits, the colour of the data  and the model} changes from blue to red over the observational baseline to visualise the non-Keplerian orbit due to the planet-planet interaction noticeable as phase shift. A similar figure, but separated into observing seasons is shown in Fig.\,\ref{fig:apsidal_precission}.
    \label{fig:RVplot_phase}
\end{figure}

\begin{figure}
    \centering
    \includegraphics[width=0.50\textwidth]{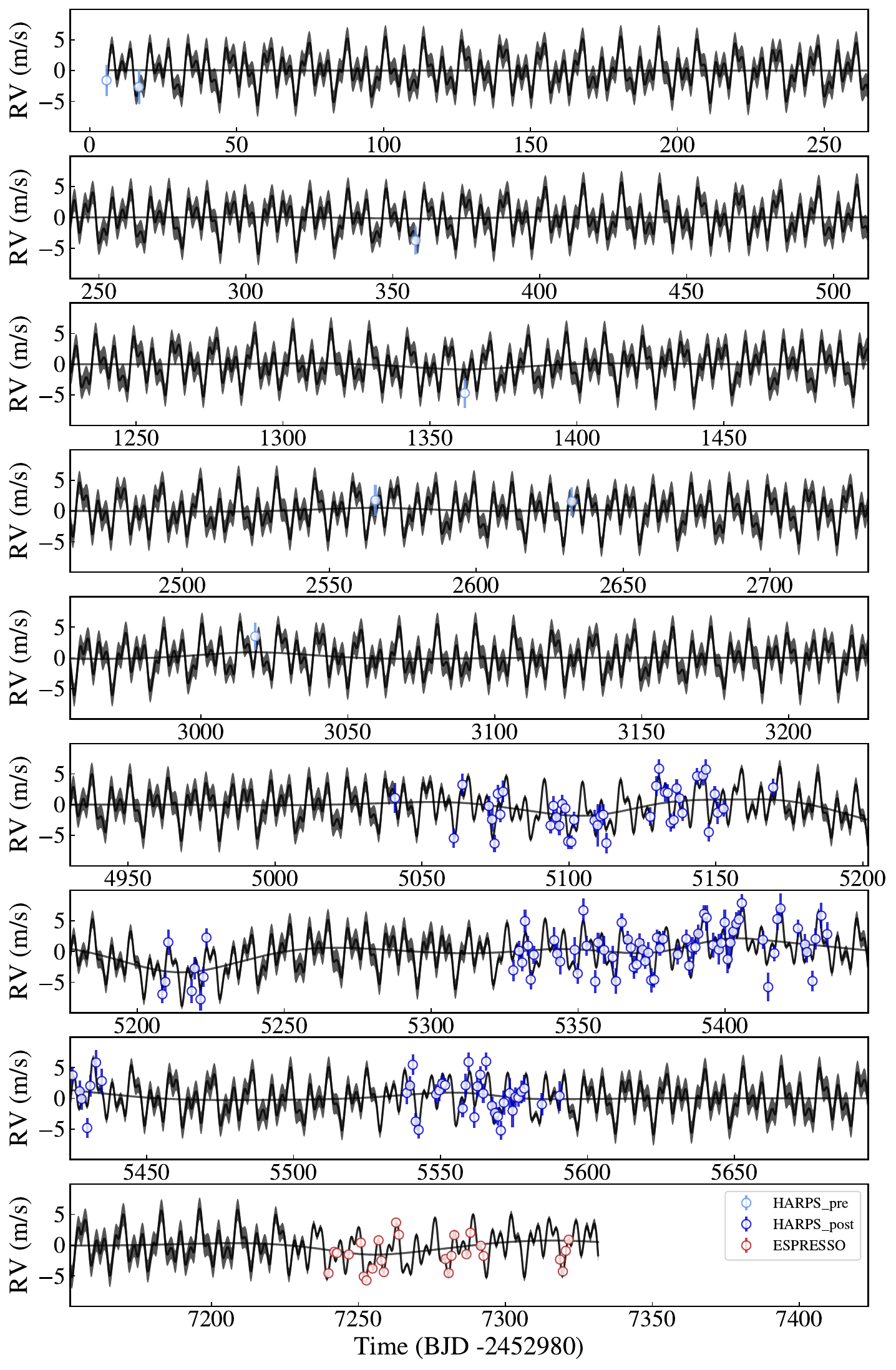}
    \caption{Data versus Model D5 (Table\,\ref{tab:mod_comp}) as a function of time. The grey shaded region comprise the $1-\sigma$ uncertainty of the model, the slowly varying line is the contribution from the rotational modulation.}
    \label{fig:RVplot}
\end{figure}

Our final best model is the co-planar three-planet model, including an {\em SHO} kernel to model the rotation induced RV variability  (D4 and D5 in Table\,\ref{tab:mod_comp}). The two models differ in the priors for the rotation period (normal versus uniform, see Table\,\ref{tab:mod_comp}) and show nearly equal values for $\ln\mathcal Z$ and we select model D5 with the wider uniform prior as the final best model. The posteriors distribution and the best parameters are those in the third and fourth column of Table\,\ref{tab:mod_dyn}. They are also displayed as corner-plots in Fig.\,\ref{fig:corner}. The main result is the determination of the  masses of the planets due to the planet-planet gravitational interaction. This is clearly visible in the phase folded RV data overlaid with the model shown in Fig.\,\ref{fig:RVplot_phase}, where we can see the orbital precession and the resulting phase shift, especially in the two inner planets. The comparison of the model including the $1-\sigma$ uncertainties with the RV data as function of time is shown in Fig.\,\ref{fig:RVplot}. The very good match visible between the data and model provides confidence in the mass determination from the planet-planet interaction. This is possible due to the $\sim$20-year baseline of HARPS data and high-precision RV measurements of the ESPRESSO data. 

Using the posterior parameter distribution, we used {\tt SPOCK} to derive a probability for each configuration to be stable. The distribution of the stability probability of Model D5 has a median of $0.74\pm 0.1$, 99.5\% are above the stability threshold used by \cite{2020PNAS..11718194T}. The probabilities for all parameter configurations are then taken as weights and stability corrected posteriors were then obtained. The parameter distributions are unaffected by the stability correction. We therefore conclude that the derived planetary architecture of the GJ\,1061 system is dynamically stable.

Recently, \citet{2022Sci...377.1211L} reported that planetary density and not radius is the optimal parameter to characterise the internal composition of sub-Neptunian planets. Using their results, showing all planets of M stars below 2\,$M_\oplus$ having Earth density, we can infer that the three planets orbiting GJ\,1061 are very likely rocky planets as opposed to water-rich or gas-rich. Assuming an Earth bulk density, the planetary radii can be inferred and are listed in Table\,\ref{tab:mod_dyn}. GJ\,1061 c and d are therefore likely rocky planets in the optimistic and conservative habitable zone as derived by \citep{2014ApJ...787L..29K}. It makes GJ\,1061 c and  d the third and fourth-closest rocky habitable zone planets, after Proxima\,b and Ross\,128\,b. 

 These results make the GJ\,1061 system an optimal target for further characterisation instruments combining high-resolution spectroscopy  with Adaptive Optics assisted high spatial resolution like the ArmazoNes high Dispersion Echelle Spectrograph \citep{2023arXiv231117075P} or the Planetary Camera Spectrograph \citep{2021Msngr.182...38K} and with space missions such as the {\em Habitable Worlds Observatory} ({\em HWO}) or the complementary {\em Large Interferometer For Exoplanets} ({\em LIFE}) mission \citep{2022A&A...664A..21Q,2022ExA....54.1197Q} in thermal emission. For the latter mission, the planets of GJ\,1061 are three out of 212 planets detectable (signal-to-noise > 7 in less than 100 h) with the reference configuration of {\em LIFE} for host stars within 20\,pc \citep{2023A&A...678A..96C}.

\begin{table*}
    \centering
    \caption{Fitted and derived planetary parameters (the median, the 16\% and 84\% percentile as well as the best value) for three Newtonian models.}
    \begin{tabular}{lc|r@{}lc|r@{}lc|r@{}lc}
         fit parameter        & prior & \multicolumn{2}{c}{posterior} & best& \multicolumn{2}{c}{posterior} & best& \multicolumn{2}{c}{posterior} & best\\
                              &                            & \multicolumn{3}{c|}{coplanar, Model D5} & \multicolumn{3}{c|}{3D} & \multicolumn{3}{c}{HARPS only} \\
         \hline
         \multicolumn{11}{c}{planet b} \\
         $P$ [d]                & $\mathcal{U}(3.206,3.209)$&3.2073 & $^{+0.0002}_{-0.0003}$ &3.2071&3.2075& $^{+0.0001}_{-0.0001}$ &3.2077&3.2066& $^{+0.0002}_{-0.0003}$ & 3.2066\\
         $m$ [$M_\oplus$]       & $\mathcal{U}(0,3)$        &1.11   & $^{+0.11}_{-0.09}$ $^{+0.15}_{-0.11}$      & 1.08  &1.11   & $^{+0.12}_{-0.11}$     & 1.09   &1.49  & $^{+0.30}_{-0.25}$     & 1.71 \\
         $h^{\prime}$                    & $\mathcal{U}(-0.4,0.4)$   &0.15   & $^{+0.11}_{-0.13}$     & 0.24  &0.10   & $^{+0.09}_{-0.14}$     & -0.17  &0.21  & $^{+0.07}_{-0.22}$     & 0.25\\
         $k^{\prime}$                    & $\mathcal{U}(-0.4,0.4)$   &-0.13  & $^{+0.10}_{-0.09}$     & 0.16  &0.15   & $^{+0.10}_{-0.09}$     &  0.13  &-0.11 & $^{+0.14}_{-0.08}$     & -0.20\\
         $\lambda$ [$^\circ$] & $\mathcal{U}(0,360)$      &229    & $^{+23}_{-24}$         & 232   &214    & $^{+21}_{-19}$         & 236    &259   & $^{+40}_{-34}$         & 246 \\
         $i$  [$^\circ$]        & $\mathcal{U}(\arccos{0},\arccos{\pi}$)& 77 & $^{+9}_{-13}$ & 70    & 65    & $^{+8}_{-7}$           & 65     &53    & $^{+9}_{-6}$           & 47\\
         $\Omega$             & $\mathcal{U}(0,360)$      & \multicolumn{3}{c|}{0 (fixed)}          &\multicolumn{3}{c|}{0 (fixed)}   & \multicolumn{3}{c}{0 (fixed)}\\[0.3cm]
         $K$ [m\,s$^{-1}$]             & derived                   &1.83    & $^{+0.18}_{-0.17}$    & 1.81  &1.78   & $^{+0.18}_{-0.17}$     &1.76    &2.18   & $^{+0.30}_{-0.30}$    &2.24\\
         $a$ [au]               & derived                   &0.0210& $^{+0.0006}_{-0.0006}$  &\,0.0210&0.0210& $^{+0.0006}_{-0.0006}$ &0.0210&0.0210& $^{+0.0006}_{-0.0006}$ &0.0210\\
         $e$                    & derived                   &0.05  & $^{+0.5}_{-0.03}$       & 0.08  &0.05   & $^{+0.4}_{-0.03}$      & 0.05   & 0.07 & $^{+0.03}_{-0.04}$ & 0.10 \\
         $r$ [$R_\oplus$]$^a$   & derived                   & 1.02& $^{+0.04}_{-0.03}$       &1.03   &1.04   & $^{+0.04}_{-0.04}$     & 1.03 & 1.14 & $^{+0.08}_{-0.07}$ & 1.20 \\
         \hline
         \multicolumn{11}{c}{planet c} \\
         $P$ [d]                & $\mathcal{U}(6.675,6.690)$& 6.6821& $^{+0.0008}_{-0.0008}$ &6.6827&6.6813& $^{+0.0004}_{-0.0004}$&6.6810&6.6830& $^{+0.002}_{-0.001}$&6.6831\\
         $m$ [$M_\oplus$]       & $\mathcal{U}(0,3)$        &1.81    & $^{+0.13}_{-0.11}$  $^{+0.21}_{-0.14}$    & 1.74  &1.81   & $^{+0.12}_{-0.11}$    & 1.88   & 2.20 & $^{+0.20}_{-0.37}$  &2.32\\
         $h^{\prime}$                    & $\mathcal{U}(-0.4,0.4)$   &-0.06   & $^{+0.11}_{-0.09}$    & -0.04 &0.05   & $^{+0.08}_{-0.08}$    & 0.11   & -0.18& $^{+0.48}_{-0.05}$  &-0.10\\
         $k^{\prime}$                    & $\mathcal{U}(-0.4,0.4)$   &-0.08   & $^{+0.10}_{-0.09}$    & -0.14 &0.00   & $^{+0.11}_{-0.10}$    & 0.13   & 0.00 & $^{+0.16}_{-0.108}$ & 0.04\\
         $\lambda$ [$^\circ$] & $\mathcal{U}(0,360)$      &337     & $^{+16}_{-16}$        & 348   &335     & $^{+17}_{-17}$       & 364    & 103  & $^{+58}_{-17}$      & 112\\
         $i$  [$^\circ$]        & $\mathcal{U}(\arccos{0},\arccos{\pi}$)& \multicolumn{3}{c|}{co-planar}& 69    & $^{+7}_{-7}$        & 60     &\multicolumn{3}{c}{co-planar}\\
         $\Omega$             & $\mathcal{U}(0,360)$      & \multicolumn{3}{c|}{0 (fixed)}           &278    & $^{+55}_{-55}$       & 217    & \multicolumn{3}{c}{0 (fixed)}\\[0.3cm]
         $K$ [m\,s$^{-1}$]             & derived                   &2.32    & $^{+0.22}_{-0.19}$   & 2.29   &2.34    & $^{+0.19}_{-0.18}$   & 2.28   &2.23  & $^{+0.26}_{-0.37}$   &2.37\\
         $a$ [au]               & derived                   &0.0342& $^{+0.0009}_{-0.0010}$ &0.0342&0.0342& $^{+0.0009}_{-0.0010}$ &0.0342&0.0342&$^{+0.0009}_{-0.0010}$&0.0342\\
         $e$                   & derived                   &0.02  & $^{+0.03}_{-0.02}$     & 0.02   &0.01  & $^{+0.01}_{-0.01}$     & 0.03   & 0.04  & $^{+0.04}_{-0.02}$  &0.01\\
         $r$ [$R_\oplus$]$^a$   & derived                   & 1.20& $^{+0.03}_{-0.03}$       &1.20   &1.22   & $^{+0.03}_{-0.03}$     & 1.23 & 1.30 & $^{+0.04}_{-0.08}$ & 1.20 \\
         \hline
         \multicolumn{11}{c}{planet d} \\
         $P$ [d]                & $\mathcal{U}(13.052, 13.075)$& 13.066& $^{+0.002}_{-0.002}$ &13.065& 13.066& $^{+0.002}_{-0.002}$ &13.068   & 13.067& $^{+0.002}_{-0.002}$&13.066\\
         $m$ [$M_\oplus$]       & $\mathcal{U}(0,3)$        &1.67    & $^{+0.17}_{-0.16}$  $^{+0.31}_{-0.21}$     & 1.69 &1.65    & $^{+0.15}_{-0.16}$  & 1.53    &1.98   & $^{+0.19}_{-0.18}$  & 2.13\\
         $h^{\prime}$                    & $\mathcal{U}(-0.4,0.4)$   &0.08    & $^{+0.12}_{-0.14}$     & -0.11&0.07    & $^{+0.10}_{-0.11}$  &-0.12    &-0.10  & $^{+0.05}_{-0.06}$  &-0.19\\
         $k^{\prime}$                    & $\mathcal{U}(-0.4,0.4)$   &-0.11    & $^{+0.15}_{-0.11}$    & 0.06 &-0.12   & $^{+0.10}_{-0.09}$  &-0.24    & -0.19 & $^{+0.14}_{-0.10}$  &-0.02\\
         $\lambda$ [$^\circ$] & $\mathcal{U}(0,360)$      &83   & $^{+26}_{-29}$            & 80   &86      & $^{+18}_{-16}$      & 86      &91     & $^{+24}_{-18}$      & 102\\
         $i$  [$^\circ$]        & $\mathcal{U}(\arccos{0},\arccos{\pi}$)& \multicolumn{3}{c|}{co-planar}& 63    & $^{+8}_{-9}$       & 61      & \multicolumn{3}{c}{co-planar}\\
         $\Omega$             & $\mathcal{U}(0,360)$      & \multicolumn{3}{c|}{0 (fixed)}          &211     & $^{+52}_{-61}$      & 328     & \multicolumn{3}{c}{0 (fixed)}\\[0.3cm]
         $K$ [m\,s$^{-1}$]             & derived                   &1.63    & $^{+0.17}_{-0.16}$     & 1.77 &1.63    & $^{+0.22}_{-0.21}$  & 1.49    &1.63    & $^{+0.18}_{-0.18}$  &1.74\\
         $a$ [au]               & derived                   &0.054& $^{+0.002}_{-0.002}$      & 0.054&0.054   & $^{+0.002}_{-0.002}$& 0.054   &0.054   & $^{+0.002}_{-0.002}$& 0.054 \\
         $e$                    & derived                   &0.04  & $^{+0.04}_{-0.03}$       & 0.02 &0.03    & $^{+0.04}_{-0.01}$  & 0.07    & 0.05    & $^{+0.04}_{-0.03}$ & 0.04\\
         $r$ [$R_\oplus$]$^a$   & derived                   & 1.16& $^{+0.04}_{-0.04}$       &1.20   &1.18   & $^{+0.04}_{-0.04}$   & 1.15   & 1.26 & $^{+0.04}_{-0.04}$ & 1.29 \\
         \hline
    \end{tabular}
    \tablefoot{A co-planar configuration where the $\Omega=0$ and the inclination of planes c and d are fixed to the one of planet b, a  three-dimensional model with $\Omega$ of the inner planet fixed due to symmetry along the light of sight, and a co-planar model using HARPS data only. In the column for model D5 we list a second set of uncertainties for the planetary masses, which are obtained from the weighted averages of the models D4-D6 in order to account for the uncertainties in the treatment of the stellar noise.\\
    \tablefoottext{a}{ The radius is estimated assuming Earth density.}}
    \label{tab:mod_dyn}
\end{table*}

\section{Summary}
We have validated the rocky nature of the three planets orbiting the nearby low-mass star GJ,1061 using 198 RV measurements collected with HARPS and ESPRESSO. The dense coverage, in part due to the RedDots campaigns (PI Jeffers, ESO ID 0101.C-0516(A) and 0112.C-2112(A)), combined with the long observational baseline and the exceptional precision of the ESPRESSO data, allowed us to detect planet-planet gravitational interactions. By modeling these interactions with N-body integrated orbits, we achieved a more comprehensive and accurate characterization of the planetary system.

We determined the  masses of all three planets orbiting GJ,1061 with a precision of about 10\%. Each of the planets has a mass below 2,$M_\oplus$, indicating that they are likely rocky planets with Earth-like bulk densities. The mutual inclinations suggest a co-planar configuration. Such well-characterized, nearby planetary systems are excellent candidates for future direct atmospheric investigations.

Additionally, using photometric time series and spectroscopic activity indicators, we determined the stellar rotation period to be approximately 125\,days. Notably, two of the planets, planets c and d, are the third and fourth-closest rocky planets to our Solar System that orbit within the optimistic habitable zone of their host star. Continued long-term monitoring, supported by high-precision RV measurements from instruments like ESPRESSO, would enable the determination of dynamical masses in more compact planetary systems orbiting M-type stars.

\begin{acknowledgements}
S.D. acknowledges support from the Deutsche Forschungsgemeinschaft under Research Unit FOR2544 ``Blue Planets around Red Stars'', project no. DR 281/39-1. SD, SVJ, FL and YT acknowledge the support of the German Science Foundation (DFG) priority program SPP 1992 `Exploring the Diversity of Extrasolar Planets' (DR 281/37.1, JE 701/5-1, TS 356/3-1). CAH and JRB acknowledge support from STFC under consolidated grant ST/X001164/1; HRAJ acknowledges support from STFC  consolidated grant ST/R000905/1. ER acknowledges financial support from the Spanish Agencia Estatal de Investigaci\'on of the Ministerio de Ciencia e Innovación through projects PID2019-107061GB-C64, PID2022-137241NB-C43, and the Centre of Excellence "Severo Ochoa" Instituto de Astrofísica de Andalucía (grant CEX2021-001131-S funded by MCIN/AEI/10.13039/501100011033).

This paper makes use of data from the MEarth Project, which is a collaboration between Harvard University and the Smithsonian Astrophysical Observatory. The MEarth Project acknowledges funding from the David and Lucile Packard Fellowship for Science and Engineering, the National Science Foundation under grants AST-0807690, AST-1109468, AST-1616624 and AST-1004488 (Alan T. Waterman Award), the National Aeronautics and Space Administration under Grant No. 80NSSC18K0476 issued through the XRP Program, and the John Templeton Foundation.

ASAS-SN is funded in part by the Alfred P. Sloan Foundation under grant G202114192. Making ASAS-SN light curves public is primarily funded by grants GBMF5490 and GBMF10501

This work has made use of data from the European Space Agency (ESA) mission
{\it Gaia} (\url{https://www.cosmos.esa.int/gaia}), processed by the {\it Gaia}
Data Processing and Analysis Consortium (DPAC,
\url{https://www.cosmos.esa.int/web/gaia/dpac/consortium}). Funding for the DPAC
has been provided by national institutions, in particular the institutions
participating in the {\it Gaia} Multilateral Agreement.

This work has made use of data from the Asteroid Terrestrial-impact Last Alert System (ATLAS) project. The Asteroid Terrestrial-impact Last Alert System (ATLAS) project is primarily funded to search for near earth asteroids through NASA grants NN12AR55G, 80NSSC18K0284, and 80NSSC18K1575; byproducts of the NEO search include images and catalogs from the survey area. This work was partially funded by Kepler/K2 grant J1944/80NSSC19K0112 and HST GO-15889, and STFC grants ST/T000198/1 and ST/S006109/1. The ATLAS science products have been made possible through the contributions of the University of Hawaii Institute for Astronomy, the Queen’s University Belfast, the Space Telescope Science Institute, the South African Astronomical Observatory, and The Millennium Institute of Astrophysics (MAS), Chile.

\end{acknowledgements}

\bibliographystyle{aa}
\bibliography{GJ1061}

\begin{appendix}
\onecolumn

\section{ESPRESSO RV data}

In Table\,\ref{tab:espresso_rv} we list the radial velocity measurements from ESPRESSO using {\tt serval}, described in Section\,\ref{sect:spec_data}. 
\begin{table}[!h]
    \centering
    \caption{RV measurements from ESPRESSO (0112.C-2112(A)).}
    \begin{tabular}{ccc}
        BJD [d] & RV [m/s] & RV error [m/s] \\
        \hline
        2460219.807560	&-3.19	&0.25\\
        2460221.745203	&0.28	&0.20\\
        2460222.708672	&0.09	&0.23\\
        2460226.705382	&-0.14	&0.28\\
        2460230.810344	&1.78	&0.22\\
        2460231.825679	&-3.73	&0.39\\
        2460232.751602	&-4.36	&0.20\\
        2460234.771253	&-2.43	&0.19\\
        2460236.852844	&2.16	&0.26\\
        2460237.848172	&-1.15	&0.23\\
        2460238.666073	&-3.00	&0.25\\
        2460242.801023	&5.04	&0.23\\
        2460243.777991	&3.06	&0.22\\
        2460259.611585	&-0.90	&0.25\\
        2460260.715458	&-3.16	&0.23\\
        2460261.666437	&-0.36	&0.31\\
        2460262.626243	&3.01	&0.28\\
        2460266.750041	&-0.10	&0.22\\
        2460267.658058	&1.49	&0.21\\
        2460268.589569	&5.30	&0.25\\
        2460271.632952	&1.28	&0.30\\
        2460272.575315	&-0.38	&0.27\\
        2460298.592932	&-0.96	&0.28\\
        2460299.562949	&-2.90	&0.28\\
        2460300.575416	&0.45	&0.24\\
        2460301.566125	&2.26	&0.26\\
        \hline
    \end{tabular}
    \label{tab:espresso_rv}
\end{table}

\section{Photometric monitoring and analysis of rotation period}

\begin{figure}[!ht]
    \centering
    \includegraphics[width=0.499\textwidth]{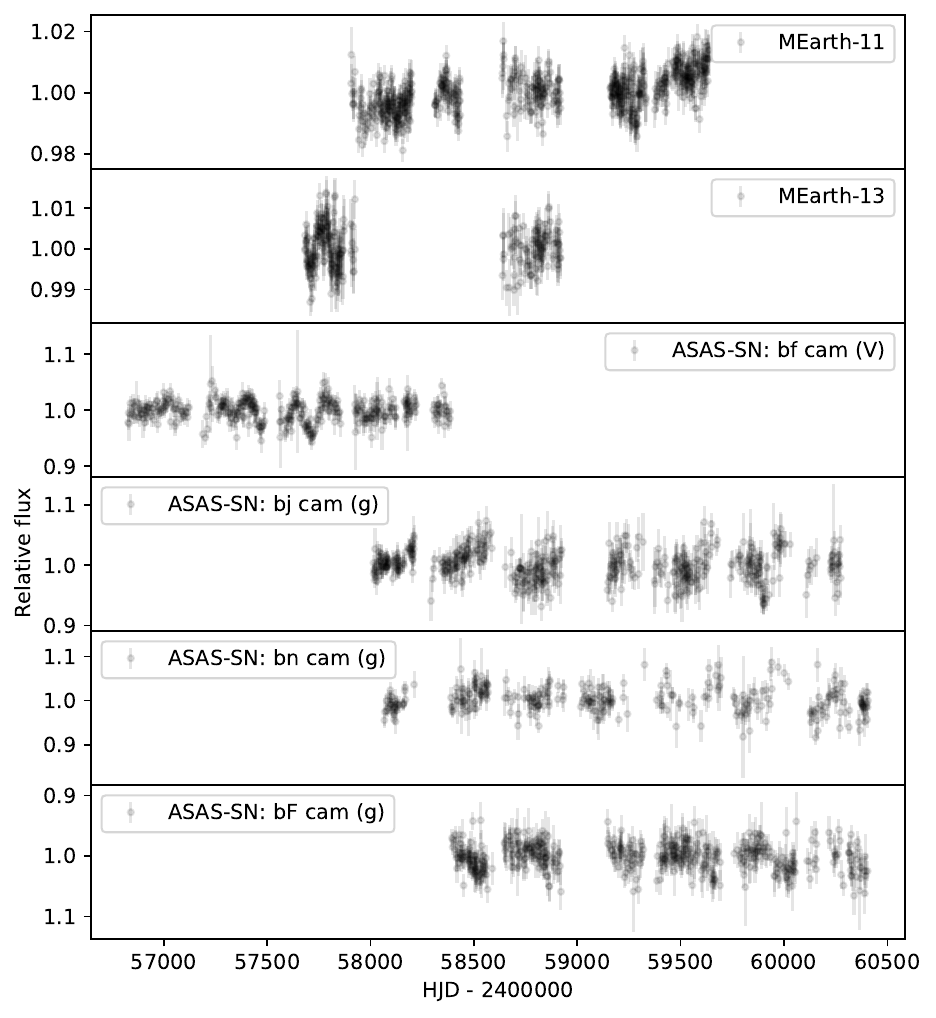}
    \caption{Time series of nightly binned photometric monitoring of GJ\,1061.}
    \label{fig:phot_timeseries}
\end{figure}

\begin{figure}
    \centering
    \includegraphics[width=0.499\textwidth]{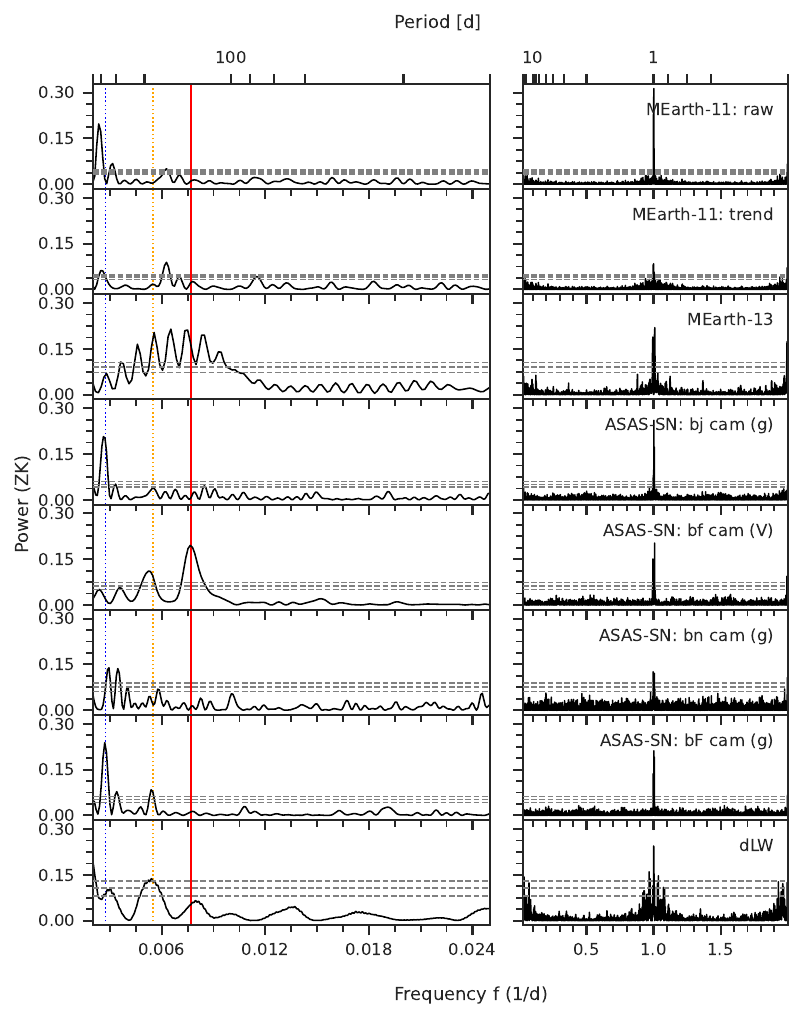}
    \caption{Left: GLS periodograms of photometric time series and the spectroscopic activity indicator differential line width (dLW). Other activity indicators do not show significant signals. The blue and yellow line mark the year and half year period. The red line indicates the potential rotation period at 125\,d. Right: the window function of the time series.}
    \label{fig:GLS_all_GJ1061_paper}
\end{figure}

The information about the instruments as well as the observing campaigns of the photometric monitoring of GJ\,1061 described in Section\,\ref{sec:phot_data} is listed in Table\,\ref{tab:phot-data}, and the time series and corresponding periodograms are displayed in Fig.\,\ref{fig:phot_timeseries} \ref{fig:GLS_all_GJ1061_paper} respectively.

\twocolumn

\begin{table}[!ht]
    \centering
    \caption{Posteriors from the best GP fit on the photometric stellar rotation period.}
    \label{tab:posteriors_Prot}
    \begin{tabular}{c c c }
        \hline
        \hline
        \noalign{\smallskip}
        Parameter                                   & Posterior                                       & Unit      \\
        \noalign{\smallskip}
        \hline
        \noalign{\smallskip}
        $P_\text{GP}$                               & $130^{+5}_{-14}$                       & d                      \\
          \noalign{\smallskip}
        $\sigma_\text{GP, MEarth-11}$               & $0.004^{+0.001}_{-0.001}$                       & ppt \\
          \noalign{\smallskip}
        $\sigma_\text{GP, bf}$                      & $0.012^{+0.003}_{-0.002}$                       & ppt \\
          \noalign{\smallskip}
        $f_\text{GP}$                               & $0.78^{+0.15}_{-0.24}$                          & \dots                  \\
          \noalign{\smallskip}
        $Q_{0, \text{GP}}$                          & $0.13^{+0.06}_{-0.03}$                          & \dots                  \\
          \noalign{\smallskip}
        $dQ_\text{GP}$                              & $16.59^{+1509.35}_{-16.36} $                    & \dots                  \\
          \noalign{\smallskip}

        $\gamma_\text{MEarth-11}$                   & $ 0.0003^{+0.0006}_{-0.0006}$                   & ppt \\
          \noalign{\smallskip}
        $\gamma_\text{bf}$                          & $0.998^{+0.002}_{-0.002}$                       & ppt \\
          \noalign{\smallskip}
        $\sigma_\text{MEarth-11}$                   & $0.0008^{+0.0006}_{-0.0005}$                    & ppt \\  
          \noalign{\smallskip}
        $\sigma_\text{bf}$                          & $0.00099^{+0.00099}_{-0.00068}$                 & ppt \\
          \noalign{\smallskip}
        intercept$_\text{MEarth-11}$                & $1.63^{+0.17}_{-0.05}$                          & \dots                  \\
        slope$_\text{MEarth-11}$                    & $\num{4.5}^{+\num{0.67}}_{-\num{0.68}}$  & \num{1e-6}                  \\
        quad$_\text{MEarth-11}$                     & $\num{2.0}^{+\num{8.0}}_{+\num{8.0}}$  & \num{1e-9}                 \\
        \noalign{\smallskip}
        \hline
    \end{tabular}
\end{table}

\onecolumn
\begin{table*}
    \caption{Overview of the photometric data. Nightly binned data have been use in the analysis.}
    \label{tab:phot-data}
    \centering
    \begin{tabular}{p{2.5cm}ccccccc}
        \hline \hline
        \noalign{\smallskip}
        Instrument                   & \multicolumn{2}{c}{Date} & Band           & {Time span}    & {Mean error}           &  \multicolumn{1}{c}{$N_\textnormal{binned}$}       \\
                                     & Begin                    & End            &                & \multicolumn{1}{c}{[d]} & {[ppt]} &                                     &     \\
        \hline
        \noalign{\smallskip}
        MEarth-11                    &  June 2017              &  February 2022  &    &  1731     & 4.71                    & 663  \\
        MEarth-13                    &  October 2016           &  March 2020     &    &  1325     & 4.07                    & 276  \\
        ASAS-SN: bf                  &  June 2014              &  September 2018 & $V$&  1557     & 16.97                   & 397  \\  
        ASAS-SN: bj                  &  September 2017         &  November 2023  & $g$&  2260     & 22.13                   & 508  \\ 
        ASAS-SN: bn                  &  November 2017          &  April 2024     & $g$&  2339     & 23.68                   & 324  \\ 
        ASAS-SN: bF                  &  October 2018           &  April 2024     & $g$&  2010     & 21.54                   & 492 \\
        ATLAS                        & August 2017             & August 2024 & $c$ & 2556 & 2 & 444 \\
        \hline
    \end{tabular}
\end{table*}

\begin{table*}[!ht]
    \centering
    \caption{Default priors used for the GP fits on the photometric data using dSHO kernel.}
    \label{tab:priors_Prot}
    \begin{tabularx}{\hsize}{lccX}
        \hline
        \hline
        \noalign{\smallskip}
        Parameter                                       & Prior                               & Unit                   & Description \\
        \noalign{\smallskip}
        \hline
        \noalign{\smallskip}
        $P_\text{GP}$                               & $\mathcal{U}(1,200)$           & d                      & Period of the GP \\
        $\sigma_\text{GP, MEarth-11 (trend previously fitted)}$                          & $\mathcal{U}(0,50)$            & ppt & Amplitude of the GP for MEarth-11 data (trend previously fitted) \\
        $\sigma_\text{GP, ASAS-SN (bf cam, $V$ filter)}$                          & $\mathcal{U}(0,50)$            & ppt & Amplitude of the GP for ASAS-SN data (bf cam, $V$ filter) \\
        $f_\text{GP}$                               & $\mathcal{U}(0, 1)$            & \dots                  & Fractional amplitude of secondary mode \\
        $Q_{\text{GP}}$                          & $\mathcal{J}(0.1, \num{1e5})$  & \dots                  & Quality factor of secondary mode \\
        $dQ_\text{GP}$                              & $\mathcal{J}(0.1, \num{1e5})$  & \dots                  & Difference in quality factor between primary and secondary mode \\

        $\gamma_\text{MEarth-11 (trend previously fitted)}$                  & $\mathcal{U}(-10, 10)$         & ppt & MEarth-11 data (trend previously fitted) zero point                                      \\
         $\gamma_\text{ASAS-SN (bf cam, $V$ filter)}$              & $\mathcal{U}(-10, 10)$         & ppt & ASAS-SN data (bf cam, $V$ filter) zero point                                      \\
        $\sigma_\text{MEarth-11 (trend previously fitted)}$                  & $\mathcal{U}(0.0, 30)$         & ppt & MEarth-11  data (trend previously fitted) jitter added in quadrature           \\ 
         $\sigma_\text{ASAS-SN (bf cam, $V$ filter)}$                  & $\mathcal{U}(0.0, 30)$         & ppt & ASAS-SN data (bf cam, $V$ filter) jitter added in quadrature           \\ 
        \hline
         \multicolumn{4}{c}{\textit{Pre-fit to MEarth-11 data (trend)}}\\
        intercept$_\text{MEarth-11}$                & $\mathcal{U}(-100, 100)$       & \dots                  & Intercept parameter of the MEarth-11 data trend  \\
        slope$_\text{MEarth-11}$                    & $\mathcal{U}(-100, 100)$       & \dots                  & Slope parameter of the Earth-11 data trend  \\
        quad$_\text{MEarth-11}$                     & $\mathcal{U}(-2, 2)$           & \dots                  & Quadratic coefficient of the MEarth-11 data trend  \\
         $\gamma_\text{MEarth-11}$                  & $\mathcal{U}(-10, 10)$         & ppt                    & MEarth-11 data  zero point    \\
         $\sigma_\text{MEarth-11}$                  & $\mathcal{U}(0.0, 30)$         & ppt                    & MEarth-11  data jitter added in quadrature           \\ 
        \noalign{\smallskip}
        \hline
   \end{tabularx}
   \tablefoot{The prior labels $\mathcal{U}$, and $\mathcal{J}$ represent uniform and log-uniform distributions, respectively.}
\rule{0cm}{6cm}   
\end{table*}

\FloatBarrier

\begin{figure}[!h]
\section{Priors and posterior distributions}
    \centering
    \includegraphics[width=0.499\textwidth]{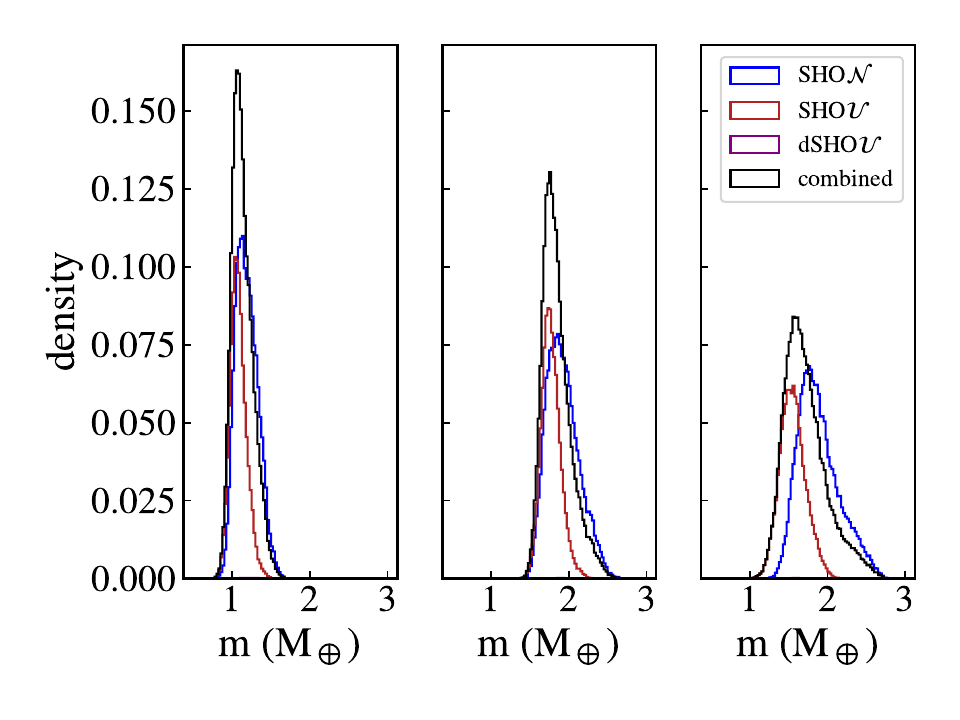}
    \caption{Posteriors for the planetary masses from models D4 to D6 using three different kernels (Table\,\ref{tab:mod_comp}) and the weighted combination of the three in black.}
    \label{fig:combined_masses}
\end{figure}

\FloatBarrier
\begin{table}[!ht]
    \centering
    \caption{Priors and posteriors for the $SHO\mathcal{U}$ kernel, RV jitter and offset for the final model.}
    \begin{tabular}{lcc}
    Hyperparameter & prior & posterior \\
    \hline
        $\sigma$ [m\,s$^{-1}$]                              & $\log \mathcal{U}$ [0.001, 10.0] & $1.63^{+0.46}_{-0.32}$\\
        P [d]                                               & $\log \mathcal{U}$ [30, 300.0]   &$124^{+28}_{-22}$\\
        $Q$                                                 & $\mathcal{U}$ [0.5, 50.0]        &$1.5^{+2.4}_{-1.2}$\\
        $\sigma_{\mathrm Jitter}$ (HARPS pre) [m\,$s^{-1}$] & $\log \mathcal{U}$ [0.05,7.5]    &$1.0^{+1.1}_{-0.6}$\\
        $\sigma_{\mathrm Jitter}$ (HARPS post) [m\,$s^{-1}$]& $\log \mathcal{U}$ [0.05,7.5]    &$0.9^{+0.2}_{-0.2}$\\
        $\sigma_{\mathrm Jitter}$ (ESPRESSO) [m\,$s^{-1}$]  & $\log \mathcal{U}$ [0.05,7.5]    &$0.6^{+0.1}_{-0.1}$\\
        offset (HARPS pre) [m\,$s^{-1}$]                    & $\mathcal{U}$ [-5,5]             &$0.7^{+1.0}_{-0.9}$\\
        offset (HARPS post) [m\,$s^{-1}$]                   & $\mathcal{U}$ [-5,5]             &$-0.1^{+0.4}_{-0.4}$\\
        offset (ESPRESSO) [m\,$s^{-1}$]                     & $\mathcal{U}$ [-5,5]             &$1.5^{+0.7}_{-0.6}$\\
        
    \end{tabular}
    \label{tab:GP}
\end{table}

\begin{figure*}
    \centering
    \includegraphics[width=0.99\textwidth]{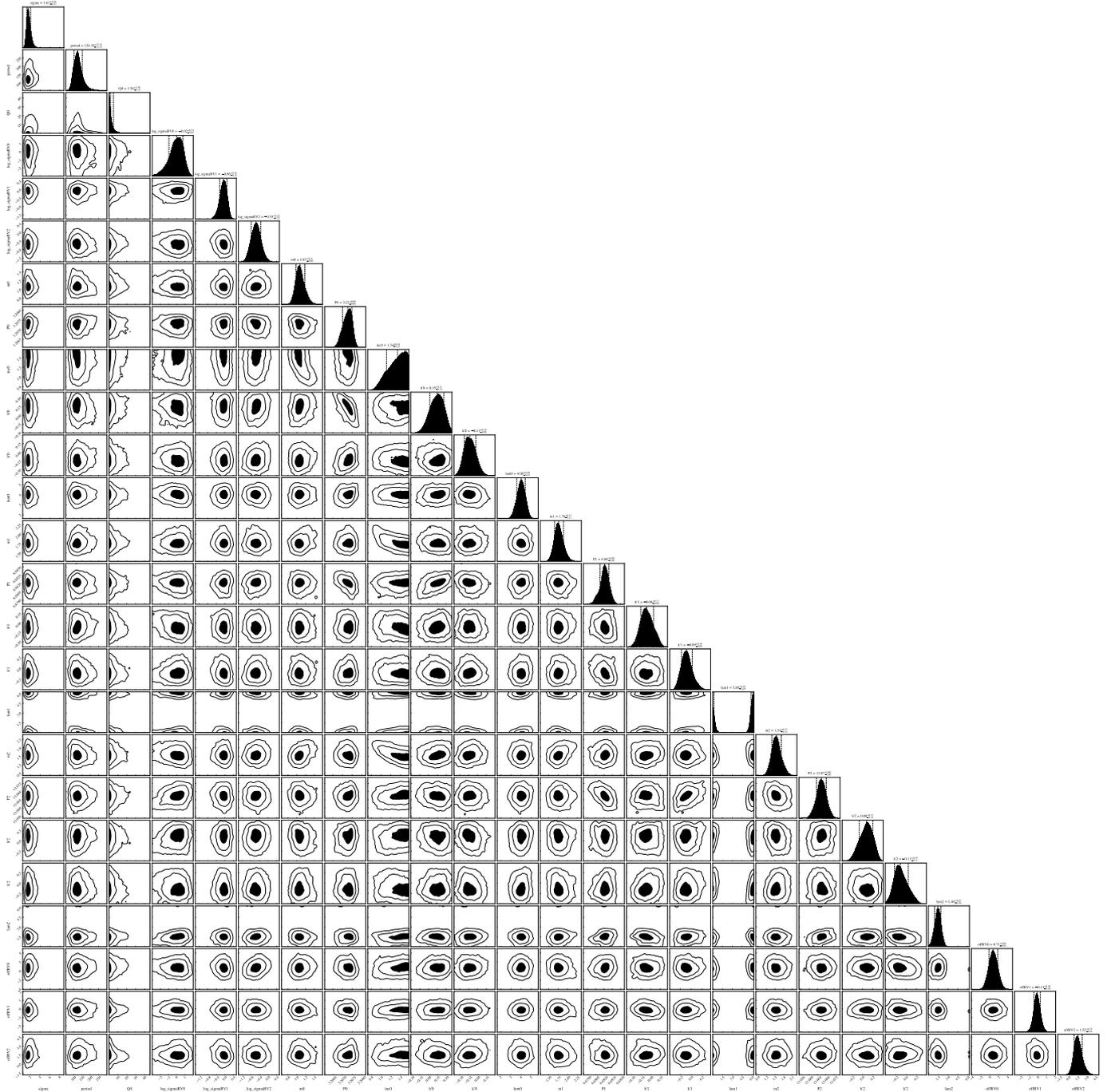}
    \caption{Corner plot  for the best-fit model D5, see Table\,\ref{tab:mod_comp}.}
    \label{fig:corner}
\end{figure*}

\FloatBarrier
\onecolumn
\section{Osculating orbital elements}
\label{sect:osculating_elements}
 In addition to Fig.\,\ref{fig:RVplot_phase}, we show here the RV data separated in seven chunks for the three planet. The non-Keplerian orbit result in a variation of the order of a minute of the orbital period, as well as in an apsidal precession. The latter we also display as a function of time for planetary system parameters randomly drawn from the posteriors.

\begin{figure*}[!ht]
    \centering
    \includegraphics[width=0.999\textwidth]{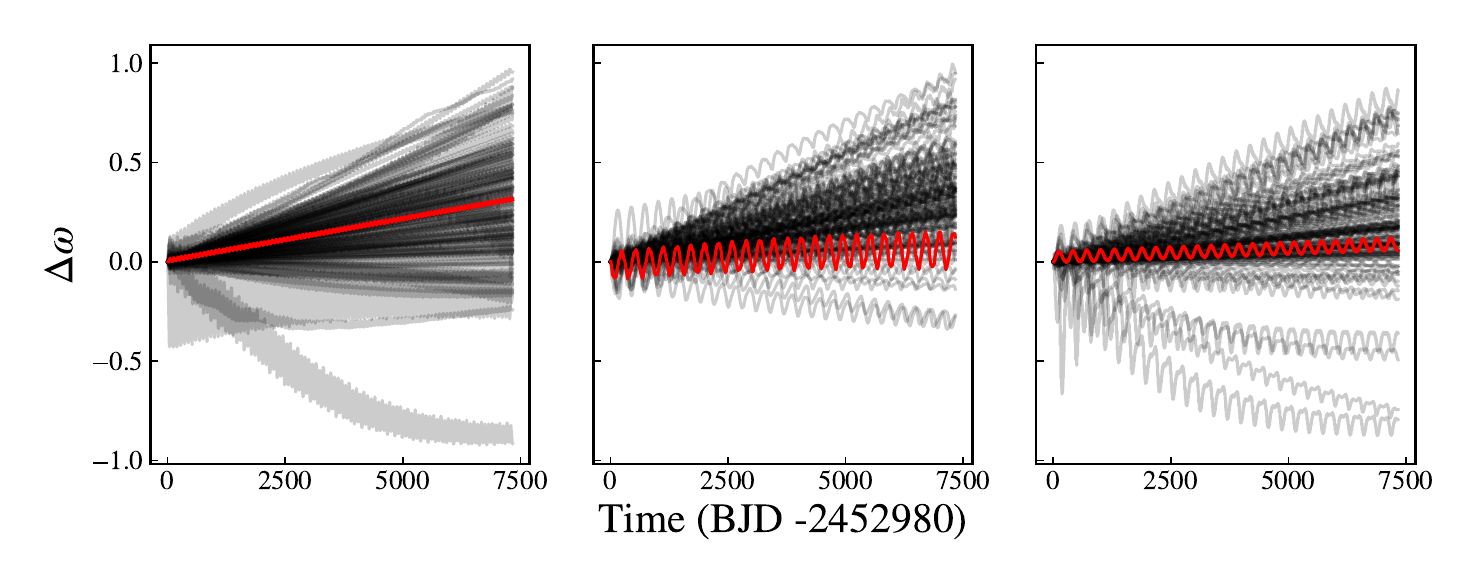}
    \caption{The change of the longitude of the pericentre (equal to the argument of pericentre in a co-planar planetary system) for planets b, c, and d from left to right. }
    \label{fig:delta_omega}
\end{figure*}

\begin{figure*}[!ht]
    \centering
    \includegraphics[width=0.99\textwidth]{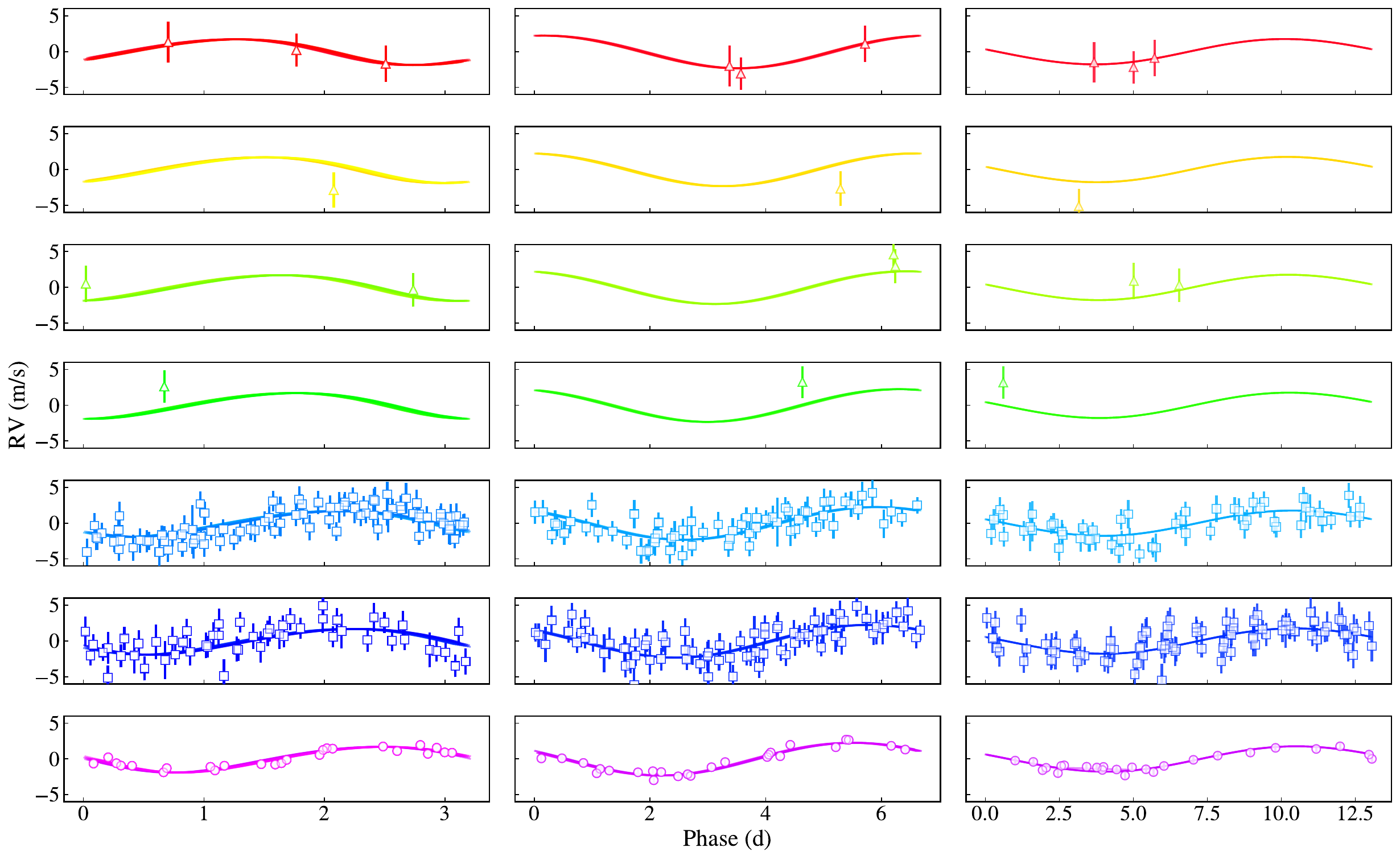}
    \caption{Alternative visualization of the non-Keplerian orbit (see Fig.\,\ref{fig:RVplot_phase}) for planets b, c, and d from left to right. The upper four panels show the spares HARPS observations overt the first $\sim 3000$ days. The next two  show the two densely sampled observing seasons with HARPS between  5000 and 5600 days, the lower panel shows the recent ESPRESSO data. }
    \label{fig:apsidal_precission}
\end{figure*}

\end{appendix}

\end{document}